\documentclass[12pt]{article}

\usepackage{soul}

\newcommand{\black}{\color{black}}
\usepackage[ruled,vlined,linesnumbered,lined,boxed,commentsnumbered]{algorithm2e}
\SetKwInput{KwInput}{Input}
\SetKwInput{KwOutput}{Output}

\usepackage{tikz}
\usetikzlibrary{arrows.meta}
\usetikzlibrary{shapes,shapes.geometric,arrows,fit,calc,positioning,automata,}

\usepackage{amsmath}
\usepackage{amsfonts}
\usepackage{relsize}
\usepackage{bm}
\usepackage{setspace}
\usepackage{float}
\usepackage{hyperref}

\definecolor{MyCol3}{rgb}{0.7,0,0}

\usetikzlibrary{shadings}
\usepackage{bbm}
\usepackage{relsize}

\usepackage{graphicx}
\usepackage[textwidth=8em,textsize=small]{todonotes}
\usepackage{amsmath}
\usepackage{natbib}

\begin{document} 

\begin{spacing}{1.3}

\begin{center}
{\large {\bf Bayesian sample size determination for multi-site replication studies}}
\end{center}
{\small
Konstantinos Bourazas$^{1}$, email: \texttt{bourazas.konstantinos@ucy.ac.cy}\\
Guido Consonni$^{2}$, email: \texttt{guido.consonni@unicatt.it}\\
Laura Deldossi$^{2}$, email: \texttt{laura.deldossi@unicatt.it}\\
$^{1}$Department of Mathematics and Statistics and KIOS Research and Innovation Center of Excellence - University of Cyprus, Nicosia, Cyprus\\
$^{2}$Department of Statistical Sciences - Universit\`{a} Cattolica del Sacro Cuore, Milan, Italy.}

\black

\begin{abstract}
An ongoing ``reproducibility crisis''  calls into question  scientific discoveries across a variety of disciplines ranging  from  life to  social sciences. Replication studies aim to investigate the validity of findings in published research,  and try to assess whether the latter are statistically consistent with those in the replications. While the majority of replication  projects are based on a single experiment, multiple independent  replications of the same experiment conducted simultaneously  
  \black 
at  different sites  are becoming more frequent. In  connection with these types of projects, we deal with  testing  heterogeneity  among sites;   specifically, we focus on  sample size determination   suitable  to deliver compelling evidence once the experimental data are gathered.

\emph{Keywords}:  analysis prior, Bayesian design, Bayes factor, design prior, heterogeneity.
\end{abstract}

\section{Introduction}

\black
Over the last few decades
it has emerged that a significant proportion of published results could not  be reproduced \citep{ioannidis2005most}, so that a
``reproducibility crisis'' \citep{baker2016reproducibility} in  empirical science has ensued. This  has been pointed out and investigated both in the social sciences, especially  psychology and economics (see among others \citealp{hensel2021reproducibility}), as well as in the life-sciences \citep{zwanenburg2019radiomics}. There are multiple reasons for this, ranging from publication bias \citep{francis2012publication} to poor experimental designs 
\citep{pashler2012replicability} and questionable statistical methodology \citep{wasserstein2016asa}. 

 \cite{simons2014value} stated that ``reproducibility is the cornerstone of science'', as it establishes the validity of  published research. Over the years there have been  numerous contributions to the analysis of replication studies;   see for instance {\cite{bonett2012replication}, \cite{zwaan2018making}, \cite{hedges2019more}}, \cite{hou2020replicating}.
 {Attempts} from the Bayesian perspective include \cite{wagenmakers2015turning}, \cite{etz2016bayesian} and
 \cite{marsman2017bayesian}. Still in the Bayesian approach,  \cite{verhagen2014bayesian} introduced the replication Bayes factor, which was implemented for fixed ANOVA designs by \cite{harms2019bayes}. Additionally,  \cite{pawel2020probabilistic} investigated replication from a predictive point of view, while  
\cite{held2020new} and \cite{held-etal-2022reverse}  exploited reverse Bayes ideas  
 to assess replication success. 
 Finally, \cite{muradchanian2021best} provided a comparative study of frequentist and Bayesian indicators for replication success.  

 Most Replication studies (RS) refer to a single replication, but interest in conducting \emph{multiple} RS   simultaneously is growing,  together with  new metrics as in  \cite{mathur2020new}. In this setting,  the analysis often  focuses on between-study variation,  especially when the issue of heterogeneity across  multiple experimental replications is of primary interest;   see  \cite{klein2014can} and  \citeauthor{hedges2021design} (\citeyear{hedges2019more}, \citeyear{hedges2019statistical} and \citeyear{hedges2021design}). In a more general framework, \cite{gronau2021primer} proposed a Bayesian model-averaged procedure to test both the presence of an effect size as well as that of the heterogeneity.

 An important aspect to underline is that very seldom do investigations offer statistical arguments regarding the choice of their  design, especially whether it was adequate to yield \emph{ex ante} compelling conclusions, e.g. in terms of power (hypothesis tests) or standard error (estimation).
Sample size determination (SSD), both in terms of the number of sites to be included in the study as well as the number of subjects within each site, is an important issue for a successful  replication design.  
\cite{wong2021design} and \cite{bonett2021design} provided reviews regarding the design and analysis of RS, while \cite{simon1999bayesian} and \citeauthor{bayarri2002a} (\citeyear{bayarri2002a} and \citeyear{bayarri2002b})  investigated the design of a replication study using  a Bayesian approach. In general, one can identify two types of  design for RS:  \emph{constrained}, where the original study is taken into consideration and the problem  is typically framed as a comparison between the original and the subsequent replication study; or  \emph{unconstrained}, where the original study is excluded from analysis. \cite{hedges2021design} proposed an unconstrained multi-site design for  SSD; see also \cite{fedorov2005design} and \cite{harden2018sample}    in the context of  clinical trials. 

In this paper we consider a Bayesian unconstrained design of multi-site
replication experiments to test the presence of heterogeneity among sites, and    provide a method for SSD using the Bayes factor (BF) as a measure of evidence \citep{kass1995bayes}. One can regard the BF as the Bayesian analogue of the frequentist likelihood ratio test for hypothesis testing,
where marginal likelihoods, as opposed to maximized likelihoods, are used. An important feature of the  BF is that it can compare any pair of hypotheses (models), and is thus not restricted to nested models. Additionally, it provides evidence for each of the two  competing hypotheses, unlike the frequentist approach, where non-significant results cannot be translated as support for the null hypothesis. For further insights into the BF and its use see \cite{dienes2014using} and \cite{hoijtink2019tutorial}. 

This paper is structured as follows. In Section \ref{sec:model} we 
present a hierarchical model which accounts for heterogeneity and identify the sub-model representing no variation; next we discuss the 
important  distinction between \emph{analysis} and \emph{design} prior
in Bayesian SSD,  derive the Bayes factor for testing heterogeneity among sites, and  show how to  approximate  its prior predictive distribution using suitable algorithms. 
 In Section \ref{sec:sample}   we introduce three categories of evidence based on the Bayes factor and the corresponding  prior probabilities, then we introduce our SSD criterion and produce  optimal sample sizes under a few scenarios chosen to allow a comparison with an alternative method. 
Finally, Section \ref{sec:conclusions} provides a short discussion,   
highlighting a few points that deserve further work.
Technical details on the derivation of the Bayes factor  are provided in the  Appendix. Sensitivity analysis and further results on alternative simulation scenarios are   available as online Supplementary material to this paper along with \texttt{R}-code.


\section{Models,   priors and Bayes factor}\label{sec:model}

In this section  we describe a hierarchical model suitable to describe heterogeneity, along with the {analysis}  adopted to derive the BF  for testing variation across sites. In addition,  we discuss the design prior required  
to obtain the  (prior)   
predictive distribution of the BF under the  competing hypotheses,  
and finally 
present  two algorithms to simulate values of the BF from its predictive distribution.

\subsection{A hierarchical model to account  for heterogeneity} \label{subsec:mod}

Consider  $m$ independent sites and let $t_j$  denote the  effect size estimator for site $j$, $j=1,2,...,m$. Adopting a meta-analytic framework, we assume that $t_j$ is approximately normally distributed, centred on the site specific effect size $\mu_j$, with variance equal to $\sigma^2_j = \sigma^2 / n_j$, where $\sigma^2$ is the  unit variance.  When the sample sizes 
are moderately large, $\sigma^2$ can be assumed known because it can be  accurately estimated. To simplify the exposition for SSD we assume a fully \emph{balanced design} (see for instance  \citealp{fedorov2005design}), wherein  the same number of subjects is enrolled in each site, so that $n_j=n, \; \forall j$. 
Independently for each $j=1, \ldots, m$, 
we consider the  hierarchical model. 
\begin{align}
\label{eq:hierarc-model}
\begin{split}
t_j | \mu_j \sim N(\mu_j, \sigma^2/n) \\
\mu_j | \mu, \tau^2  \sim N(\mu, \tau^2),
\end{split}
\end{align}
where $\mu$ is the overall mean effect size and $\tau^2$ represents heterogeneity among sites; see Figure \ref{fig:DAG} for a visualisation based on a Directed Acyclic Graph (DAG). 
We will deal with  priors for $\mu$ and $\tau^2$ in Subsection \ref{Prior}.

\begin{center}
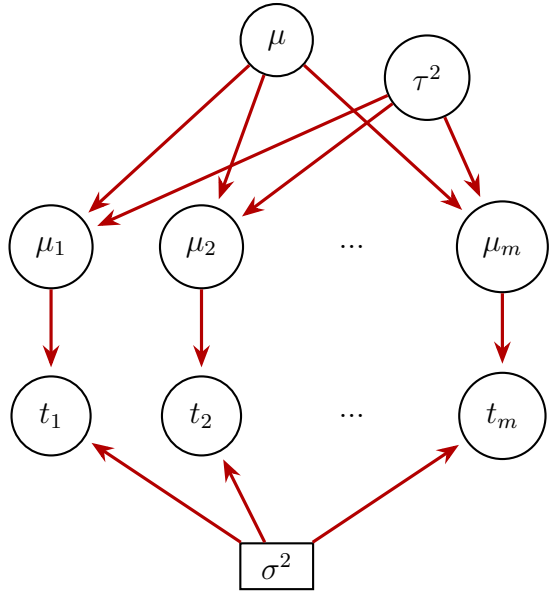
\begin{figure}[ht]
\centering

\tikzset{elliptic state/.style={draw,ellipse, thick}}
\tikzset{circle state/.style={draw,circle, thick}}
\tikzset{rectangle state/.style={draw,rectangle, thick}}
\begin{tikzpicture}[shorten >=3pt,auto, scale = 1, transform shape]

      \node[circle state] (A)  at (9,11) {$ \mbox{
 $\mu$ } $};

      \node[circle state] (B)  at (11,10.5) {$ \mbox{
 $\tau^2$ } $};

      \node[circle state] (C)  at (6,8.25) { $ \mbox{
 $\mu_1$ } $ };
      \node[circle state] (D)  at (8,8.25) { $ \mbox{
 $\mu_2$ } $ };
       \node[] (E)  at (10,8.25) {...};
       \node[circle state] (F)  at (12,8.25) { $ \mbox{
 $\mu_m$ } $ };

  \begin{scope}[>={Stealth[MyCol3]},
              every node/.style={fill=white,circle},
              every edge/.style={draw=MyCol3,very thick}]
    \path [->] (B) edge  (C);
    \path [->] (B) edge  (D);
    \path [->] (B) edge  (F);

\end{scope}

 \begin{scope}[>={Stealth[MyCol3]},
              every node/.style={fill=white,circle},
              every edge/.style={draw=MyCol3,very thick}]
    \path [->] (A) edge  (C);
    \path [->] (A) edge  (D);
    \path [->] (A) edge  (F);

\end{scope}

      \node[circle state] (H)  at (6,6) { $ \mbox{
 $t_1$ } $ };
      \node[circle state] (I)  at (8,6) { $ \mbox{
 $t_2$ } $ };
       \node[] (J)  at (10,6) {...};
       \node[circle state] (K)  at (12,6) { $ \mbox{
 $t_m$ } $ };

 \begin{scope}[>={Stealth[MyCol3]},
              every node/.style={fill=white,circle},
              every edge/.style={draw=MyCol3,very thick}]
    \path [->] (C) edge  (H);
    \path [->] (D) edge  (I);
    \path [->] (F) edge  (K);

\end{scope}


      \node[rectangle state] (G)  at (9,4) {$ \mbox{
 $\sigma^2$ } $};

  \begin{scope}[>={Stealth[MyCol3]},
  every node/.style={fill=white,circle},
              every edge/.style={draw=MyCol3,very thick}]
    \path [->] (G) edge  (H);
    \path [->] (G) edge  (I);
    \path [->] (G) edge  (K);

\end{scope}


\end{tikzpicture}

\centering
\caption{DAG of  hierarchical model \eqref{eq:hierarc-model}}
\label{fig:DAG}
\end{figure}
\end{center}
\vspace{-1cm}
Integrating out $\mu_j$ in  \eqref{eq:hierarc-model} we obtain
\begin{align}
 \label{eq:M1}
\mathcal{M}_1: t_j | \mu, \tau^2 \sim N(\mu, \tau^2 + \sigma^2/n).
\end{align}

Setting $\tau^2=0$ in \eqref{eq:M1}
 gives rise to the model of no-heterogeneity
\begin{align} 
\label{eq:M0}
\mathcal{M}_0: t_j | \mu \sim N(\mu, \sigma^2/n).
\end{align}

We will compare models $\mathcal{M}_0$ and $\mathcal{M}_1$
in Subsection \ref{subsec:BF}
using the Bayes factor (BF).

\subsection{Prior distributions} \label{Prior}

Consider first $\tau^2$.  Its plausible range of values is 
better appreciated in relation to the unit variance  $\sigma^2$.
As a consequence, and reverting to standard deviation, we  work in terms of the  \emph{relative heterogeneity}  $\gamma = \tau/\sigma$.

Recall that an 
experimental design is a \emph{prospective} enterprise and is meant to achieve a desired  level of inferential performance before the data come in. This translates to an adequate sample size that has to be determined. Unlike  frequentist power analysis, which is \emph{conditional} on a fixed value of the parameter $\gamma$ to be tested, the Bayesian approach requires a full prior on the parameter space.  
In this context it is common to distinguish between two types of priors: 
the {analysis} prior, which we label as $h_a(\gamma)$, and the {design} prior 
denoted by $h_d(\gamma)$; see \cite{o2001bayesian} and \cite{o2005assurance}, who used them in the setting of clinical trials.  The \emph{analysis} prior is used to make inference once the data come in and, in our case, it will be used to evaluate the Bayes factor. In principle, it should be weakly informative, so that it can be broadly acceptable, and 
yet proper so that the BF can be unambiguously evaluated. On the other hand, the \emph{design} prior should be an informative prior, representing the position of the researcher about the size of the heterogeneity that is expected or  is deemed interesting to detect.

We first consider the analysis prior $h_a(\gamma)$.
Based on \cite{rover2021weakly}, who provided a wide review on  weakly informative priors for heterogeneity,  we assume a  Half-$t(\nu_\gamma,\;\sigma_\gamma)$ distribution, where $\nu_\gamma$ is the degrees of freedom and $\sigma_\gamma$ is the scale parameter. The Half-$t(\nu_\gamma,\;\sigma_\gamma)$
is the distribution of the absolute value of  a Student-t variate  centered at zero with
 degrees of freedom $\nu_\gamma$  and scale $\sigma_\gamma$.
 Its density   is monotonically decreasing with heavy upper tail for small values of $\nu_\gamma$. Regarding the choice of the hyper-parameters, we set $\nu_\gamma=4$; see   \citet{rover2021weakly}. Moreover  we set $\sigma_\gamma=1/7$, so that the 95\% quantile of the Half-t distribution is about 0.4,  a value  broadly in line with the suggestions of \cite{hedges2021design} for a variety of applied domains. 

\begin{figure}[ht]
\centering
\includegraphics[width=1 \textwidth]{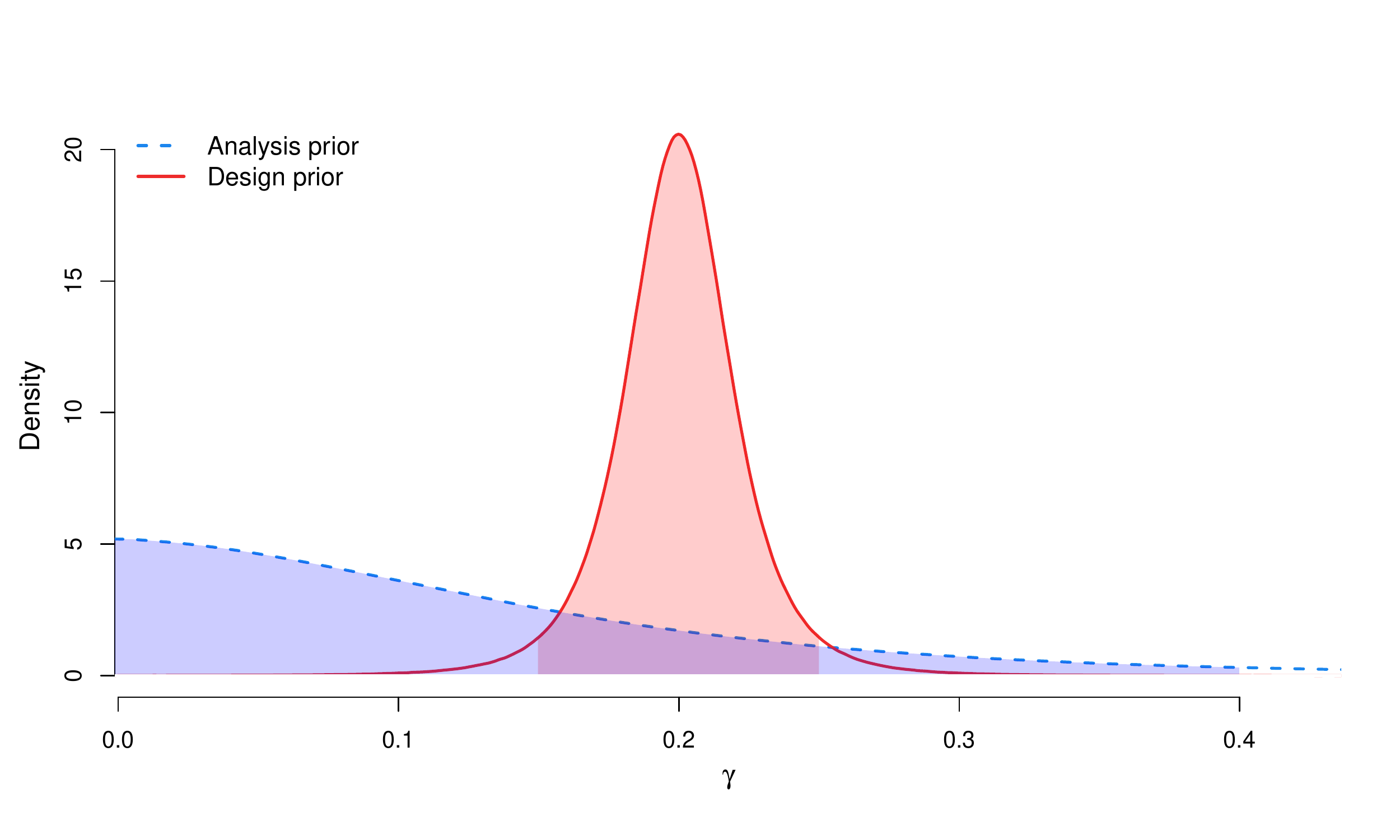}
\centering
\vspace{-1cm}
\caption{The analysis prior $h_a(\gamma)$ (dotted line) and the design prior $h_d(\gamma)$ (solid line), along with their credible interval $95\%$ (highlighted regions). }
\label{fig:analysis_design}
\end{figure}

We now address the issue of the  design prior $h_d(\gamma)$, which we take to be a Folded-t distribution \citep{psarakis1990folded}. The Folded-t is the  distribution of the absolute value of a variate having a non-standardized t-distribution with location    $\mu_\gamma$, scale $\sigma_\gamma$ and degrees of freedom $\nu_\gamma$; when  $\mu_\gamma=0$ it reduces to the Half-t.   Regarding the hyper-parameters, we let $\nu_\gamma=4$ and set $\mu_\gamma=0.2$; the latter choice  is based on the evaluation of plausible values for the { relative heterogeneity} $\gamma$    discussed in \cite{hedges2021design}. Adding a moderate amount of uncertainty around the location such as $\pm 0.05$,  we tune $\sigma_\gamma=1/55$ to achieve a credible interval of 95\% coverage for the region $[0.15, 0.25]$. Figure \ref{fig:analysis_design} provides the density plot of the analysis and the design prior used in the current set-up. Sensitivity analysis is carried out  in the Supplementary material for alternative configurations of the hyper-parameters.

Consider now the prior for $\mu$ appearing  both in  model $\mathcal{M}_1$ and $\mathcal{M}_0$; see  equations \eqref{eq:M1} and \eqref{eq:M0} respectively. The overall mean represents a nuisance parameter when testing heterogeneity, and we take it to be independent of $\gamma$ \emph{a priori}. We suggest using the Jeffreys prior  
$\pi (\mu) \propto 1$. 
Despite being improper,  it represents a suitable  choice for the derivation  of the Bayes factor in our case,  
because $\mu$ is a parameter common to both models.

\black
\subsection{ The Bayes factor and its predictive distribution} \label{subsec:BF}

We compare models $\mathcal{M}_0$ and $\mathcal{M}_1$ through the Bayes factor $BF_{01}$,  given by the ratio between the marginal data distribution under $\mathcal{M}_0$ and $\mathcal{M}_1$.  The measure  $BF_{01}$ quantifies the support for the null  over the alternative model; for instance  the value $BF_{01}=3$ states  that $\mathcal{M}_0$ is three times more likely than $\mathcal{M}_1$. \cite{Jeffreys:1961}  proposed a heuristic classification scheme to interpret the evidence provided by the BF, grouping values into a few   categories; for subsequent elaborations see \cite{kass1995bayes}  and  \cite{schonbrodt2018bayes}. 

Based on \eqref{eq:M0} and  \eqref{eq:M1} one obtains
the marginal data distribution under   model
$\mathcal{M}_0$, respectively $\mathcal{M}_1$,  
{\small
\begin{align}
m_0(\bm{t_r}) & =  \mathlarger{ \int\limits _{-\infty}^{+\infty} f(\bm{t_r} | \mu, \mathcal{M}_0) \pi(\mu) d\mu = \left(\frac{1}{m}\right)^{1/2}  \cdot \left(\frac{2\pi \sigma^2}{n}\right)^{(1-m)/2} \cdot \exp \left\{ -\frac{Q}{ 2}  \right\}, } \\
\begin{split}
m_1(\bm{t_r}) & =  \int\limits _{0}^{+\infty} \int\limits _{-\infty}^{+\infty} f(\bm{t_r} | \mu,\gamma, \mathcal{M}_1) h_a(\gamma) \pi(\mu)  d\mu d\gamma \\
& =   \left(\frac{1}{m}\right)^{1/2}  \cdot \left(2\pi \sigma^2\right)^{(1-m)/2} \int\limits _{0}^{+\infty} \left(\frac{1}{n}+\gamma^2\right)^{(1-m)/2} \exp \left\{ -\frac{Q}{ 2} \cdot  \left( 1+n\gamma^2\right)^{-1}  \right\} h_a(\gamma) d\gamma,
\end{split}
\end{align}}
where $\bm{t_r}=(t_1, \; ..., \; t_m)$, $\bar{t}$ is the sample mean of the $t_j$'s
and $Q=n \cdot \sum_{j=1}^m \left( t_j - \bar{t} \right)^2 / \sigma^2 $ is a quantity whose distribution does not depend on $\sigma^2$. It appears that $m_0(\bm{t_r})$ is available in closed form. On the other hand the evaluation of  $m_1(\bm{t_r})$ can be approximated using  Monte Carlo simulation.  The resulting BF is:
{\small
\begin{align} 
\label{eq:BF_form}
BF_{01}(\bm{t_r})  =  \dfrac{m_0(\bm{t_r})}{m_1(\bm{t_r})} = \dfrac{n^{(m-1)/2} \cdot \exp \left\{ -\dfrac{Q}{ 2}  \right\}}{\mathlarger{ \int\limits _{0}^{+\infty}} \left(\dfrac{1}{n}+\gamma^2\right)^{(1-m)/2}  \cdot \exp \left\{ -\dfrac{Q}{ 2} \cdot  \left( 1+n\gamma^2\right)^{-1}  \right\} h_a(\gamma) d\gamma},
\end{align}
}
which  depends on the data only through $Q$.  
\black 
We emphasise that,  at the design stage, the observations $\bm{t}_r$ are not yet available.
As a consequence, planning for unambiguous results in terms of the BF  requires 
 the \emph{prior predictive}  distribution of 
$BF_{01}(\bm{t_r})$, which in turn depends on the (prior) predictive distribution of $Q$.
\black 
 Two well known facts  are: 
i) under   $\mathcal{M}_0$,  
$Q \sim \chi_{m-1}^2$,      where $\chi_{p}^2$ denotes a chi-squared distribution with $p$ degrees of freedom; 
ii) under  $\mathcal{M}_1$, and  conditionally on $\gamma$, $(1+n \gamma^2)^{-1}  \cdot Q \sim \chi_{m-1}^2$  
\citep{hedges2001power}. To obtain the unconditional distribution of $Q$
under $\mathcal{M}_1$, a further mixing wrt $\gamma \sim h_d(\gamma)$ is required.
Finally, to obtain a realization of   $BF_{01}(\bm{t_r})$ from its prior  predictive distribution, items i) and ii) above  must be coupled with  the evaluation of the  integral appearing  in the denominator of \eqref{eq:BF_form}.   
\black
The above computational program will be carried out using a Monte Carlo approximation. Algorithm \ref{alg:M0} 
describes the procedure when the assumed model is $\mathcal{M}_0$, while Algorithm 
\ref{alg:M1} refers to $\mathcal{M}_1$.


\begin{algorithm}{
		\SetAlgoLined
		\vspace{0.1cm}
		\KwInput{number of sites $m$, number of subjects per site $n$, random sample of size $S$ generated from the analysis prior $\bm{\gamma}_a$, number of iterations $T$ } 
  \vspace{0.1cm}
		\KwOutput{vector $ \bm{BF}_{01}$  of size $T$ from the predictive distribution of $\textnormal{BF}_{01}$ under $\mathcal{M}_0$}
  \vspace{0.1cm}
		\For{$t=1,\dots,T$}{
  \vspace{0.1cm}
			Generate $q \sim \chi_{m-1}^2 $ \\[4pt] 

$g_0 \gets n^{(m-1)/2} \cdot \; \exp \left\lbrace -q/2 \right\rbrace$

$g_1 \gets  \dfrac{1}{S}  \sum_{j=1}^S  \left[ 1/n+\gamma_{a,j}^2\right]^{(1-m)/2 } \cdot \; \exp \left\{ -q/2 \cdot \left(1+n\gamma_{a,j}^2 \right)^{-1} \right\}$

$BF_{01,t} \gets g_0 / g_1$
\vspace{0.1cm}
		}}	
\caption{Simulating from the prior predictive distribution of $BF_{01}$ under $\mathcal{M}_0$}
\label{alg:M0}
\end{algorithm}

\begin{algorithm}[ht]{
		\SetAlgoLined
		\vspace{0.1cm}
		\KwInput{number of sites $m$, number of subjects per site $n$, random sample of size $S$ generated from the analysis prior $\bm{\gamma}_a$, number of iterations $T$, random sample of size $T$ generated from the design prior $\bm{\gamma}_d$ } 
  \vspace{0.1cm}
		\KwOutput{vector $ \bm{BF}_{01}$  of size $T$ from the predictive distribution of $\textnormal{BF}_{01}$ under $\mathcal{M}_1$}
  \vspace{0.1cm}
		\For{$t=1,\dots,T$}{
  \vspace{0.1cm}
			Generate $q \sim \chi_{m-1}^2 $ \\[4pt] 
   
$ q' \gets q' \cdot (1+n \gamma_{d,t}^2) $ \\[4pt]

$g_0' \gets n^{(m-1)/2}\; \cdot \; \exp \left\lbrace -q'/2 \right\rbrace$

$g_1' \gets  \dfrac{1}{S}  \sum_{j=1}^S  \left[ 1/n+\gamma_{a,j}^2\right]^{(1-m)/2 } \cdot \; \exp \left\{ -q'/2 \cdot \left(1+n\gamma_{a,j}^2 \right)^{-1} \right\}  $

$BF_{01,t} \gets g_0' / g_1' $
\vspace{0.1cm}		}}	
	\caption{Simulating from the prior  predictive distribution of $BF_{01}$ under $\mathcal{M}_1$}
    \label{alg:M1}
\end{algorithm}

\begin{figure}[ht]
\centering
\includegraphics[width=1 \textwidth]{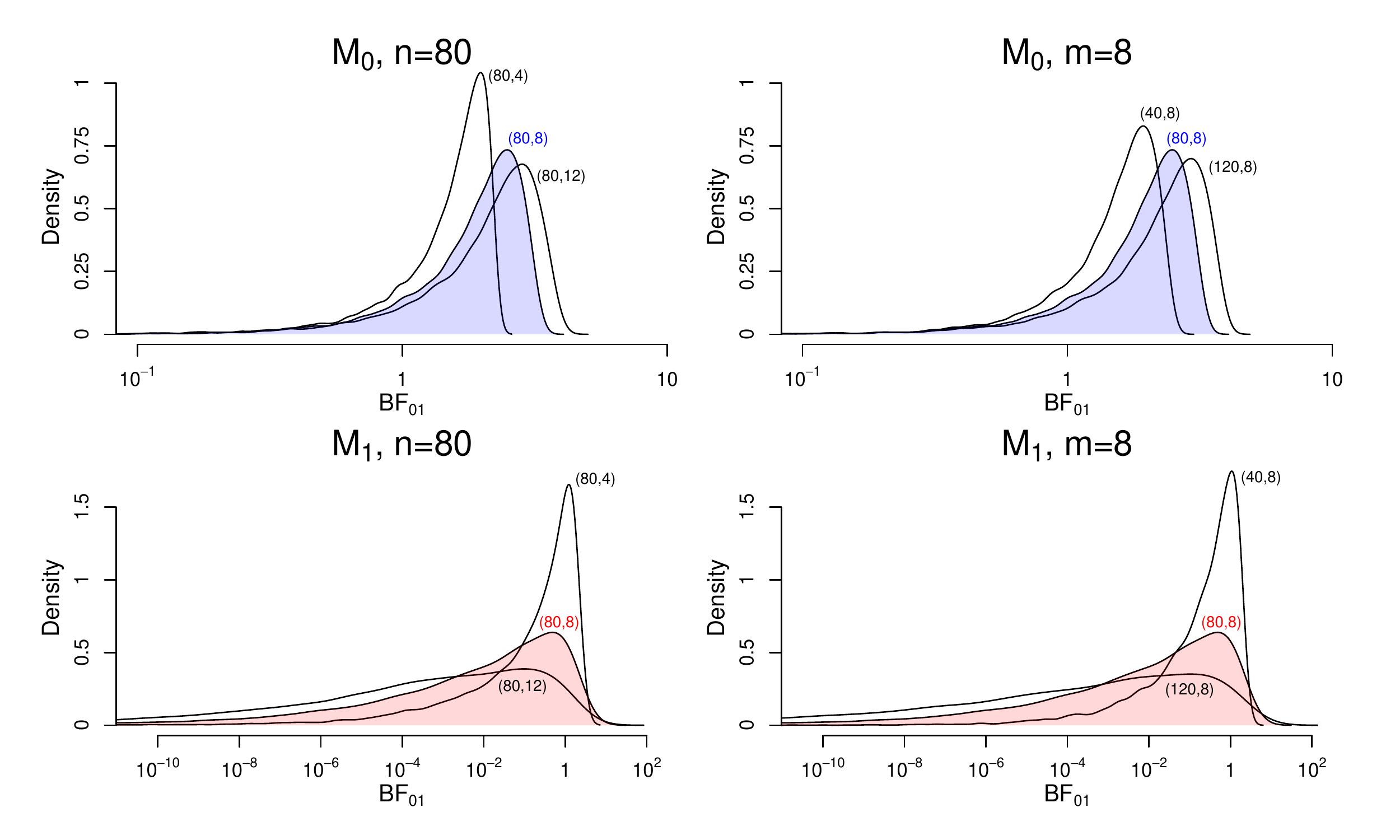}
\centering
\vspace{-1cm}
\caption{Prior predictive  distribution of BF under $\mathcal{M}_0$ (top row) and $\mathcal{M}_1$ (bottom row) for selected pairs of $(n,m)$ with  $n \in \{40,\;80,\;120 \}$ and  $m \in \{4,\;8,\;12 \}$. }
\label{fig:evolution_BF}
\end{figure}

Figure \ref{fig:evolution_BF}  reports the prior predictive  distribution of the BF (in logarithmic scale) under  model $\mathcal{M}_0$ (top panels) and $\mathcal{M}_1$ (bottom panels)  for a few pairs  $(n,m)$ whose product   $n \cdot m$ is kept constant. 
Specifically, in the two left panels we set $n=80$, with $m \in \{4,\;8,\;12 \}$; while in the right panels we fixed $m=8$ with  $n  \in \{40,\;80,\;120 \}$. 
Reading row-wise (i.e. for a fixed true model),  each distribution on the left has a corresponding distribution on the right with the same $n \cdot m$. In this way one can better appreciate the effect of reallocating a given total number of subjects between number of sites $m$ and number of  subjects per site $n$.   
For visualization purposes, the distributions for the pair $(n,m)=(80,8)$ are highlighted (in blue under $\mathcal{M}_0$ and in red under $\mathcal{M}_1$), as they are common in the two settings. 

Figure \ref{fig:evolution_BF}
reveals the strong imbalance in the learning rate of the BF  distribution under each of the two models. 
It is apparent  that under 
$\mathcal{M}_0$ the maximal $\log(BF_{01})$, representing evidence in \emph{favour} of the true model, never attains the value 10 while  under $\mathcal{M}_1$ the corresponding evidence can be as large  as $10^{6}$ or beyond   (recall that
$BF_{10}=1/BF_{01}$, where $BF_{10}$  measures evidence in favour of $\mathcal{M}_1$).  

This phenomenon has been investigated from a  theoretical perspective in  \cite{Dawid:2011},  where it is essentially shown that for nested models, as in our setup, the BF grows with the square root of the sample size under the null model, while its growth is exponential under the larger encompassing model; see also \cite{Johnson:Rossell:2010}.
This result however  remains relatively neglected in  papers,  a few notable exceptions being 
 \cite{schonbrodt2018bayes} and \citet{ly2022bayes}.

\section{Sample size determination }\label{sec:sample}

 In this section we consider probabilities of correct, misleading and  undetermined evidence when the true  model is either 
  $\mathcal{M}_0$ or $\mathcal{M}_1$.
 Based on these probabilities,  we provide a design framework   to determine   configurations for the pair  $(n,m)$ capable to deliver  compelling  evidence when testing  heterogeneity.

\subsection{
Bayes factor thresholds and classification  of model evidence}
\label{subsec:evaluation}

\begin{figure}[ht]
\centering
\includegraphics[width=1 \textwidth]{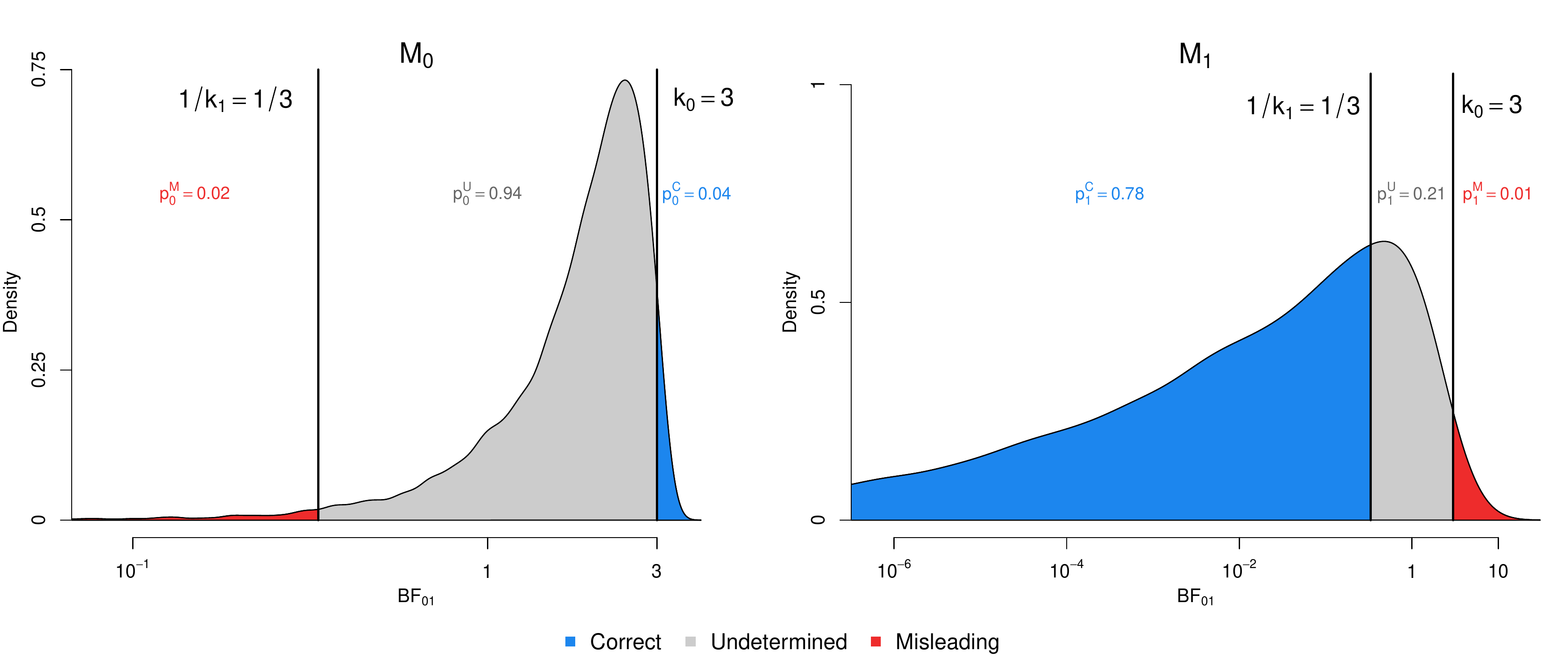}
\centering
\vspace{-1cm}
\caption{The probabilities of Correct, Misleading and Undetermined evidence under ${ \mathcal{M}_0}$ and ${ \mathcal{M}_1}$ for $(n,m)=(80,8)$. }
\label{fig:probs}
\end{figure}

For given positive  thresholds $k_0$ and $k_1$,  if $BF_{01}$ is greater than $k_0$ then  data suggest  evidence in favor of $\mathcal{M}_0$  (at level $k_0$),    while  if it is less than $1/k_1$, then data   suggest  evidence  in favor of $\mathcal{M}_1$ (at level $k_1$).   Finally,  if the BF  lies in the interval $(1/k_1, k_0)$,  evidence is undetermined.
 If $k_0$ is set at a high value such as 10 or higher, then $BF_{01}>k_0$
can be regarded as \emph{strong}  evidence in favor of $\mathcal{M}_0$ \citep{kass1995bayes,  schonbrodt2018bayes} or, in the words of \cite{De:Santis:2004},  \emph{decisive} evidence for $\mathcal{M}_0$. Similar considerations apply for $k_1$ as far as evidence for $\mathcal{M}_1$ is concerned. Smaller values of $k_0$ and $k_1$ such as 3 or 5 - which are sometimes  used  \citep{weiss1997bayesian} - would  instead only suggest  \emph{moderate} evidence in either direction. Notice that distinct thresholds $k_0$ and $k_1$ are allowed, usually with $k_0<k_1$ because learning the true model is slower under  $\mathcal{M}_0$ than under $\mathcal{M}_1$; see Figures \ref{fig:evolution_BF} and  \ref{fig:probs}, and our comments at the end of  Subsection \ref{subsec:BF}. 

Rather than fixing upfront $k_0$ and $k_1$ as evidence thresholds,  one can specify  them  \emph{indirectly} through design-based considerations such as  the probability  of Type I error $\alpha$  and  power  $(1-\beta)$ both interpreted from the Bayesian perspective. In this way the BF acts as a  mere test statistic, so that  its intrinsic meaning, along with the  substantive interpretation of the evidence cut-offs $k_0$ and  $k_1$,   are forfeited. On the other hand,  the resulting  cut-off values  satisfy more conventional design goals and thus might be more broadly acceptable to practitioners.

Given thresholds $k_0$ and $k_1$, and assuming  that either
 $\mathcal{M}_0$ or $\mathcal{M}_1$ is in turn the true
 data-generating model, we define the following events (omitting for brevity dependence on the chosen thresholds):
\begin{itemize}
\item  
\emph{Correct} evidence ($C$): 
evidence is in favour of the correct model
\item 
\emph{Misleading}  evidence ($M$): 
evidence is in favour of the incorrect model
\item 
\emph {Undetermined} ($U$) evidence: neither $C$ nor $M$ hold. 
\end{itemize}

The \emph{conditional} (prior predictive) probability of each of the above events is naturally evaluated  under the assumption that either model in turn  holds true.
This can be approximated,  separately  under $\mathcal{M}_0$ and $\mathcal{M}_1$,   applying  Algorithm \ref{alg:M0} and \ref{alg:M1}, respectively. 
 One can also evaluate the \emph{unconditional}, or \emph{overall},   probability of the events $\{C,M,U\}$ 
by averaging the corresponding conditional probabilities across the two models, using prior model probabilities 
$\pi_0=p(\mathcal{M}_0)$ and 
$\pi_1=p(\mathcal{M}_1)=1-\pi_0$; 
see Table 1 for a summary. 
 

Figure \ref{fig:probs} reports the distribution of $BF_{01}$ under
$\mathcal{M}_0$ (left panel) and  $\mathcal{M}_1$ (right panel)
highlighting the corresponding probabilities of  Correct, Misleading and Undetermined evidence   for $(n,m)=(80,8)$ and  $k_0=k_1=3$. It is apparent that the conditional  probability of Correct evidence is appreciably higher under $\mathcal{M}_1$ (78\%) than under $\mathcal{M}_0$ (4\%);    correspondingly that of Undetermined evidence is higher under $\mathcal{M}_0$
(94\%) than under $\mathcal{M}_1$ (21\%). 
This reinforces the fact that learning under   $\mathcal{M}_0$  proves to be harder.
\black 
\begin{table}[H]
\caption{The probabilities of  Correct, Misleading and Undetermined evidence.}
\label{tab:correct-misleading-undetermined}
\black
\vspace{0.2cm}
\centering {
\resizebox{\textwidth}{!}{
\begin{tabular}{lccc}
  & {\emph {Correct evidence}}  &  {\emph {Misleading evidence}} & {\emph {Undetermined evidence}}\\
  \hline
\hline \\
  ${\mathcal{M}_0}$ & $p_0^{C}=p(BF_{01}>k_0| \mathcal{M}_0)$  &  $p_0^{M}=p(BF_{01}<1/k_1| \mathcal{M}_0)$ &  $p_0^{U}=p(1/k_1<BF_{01}<k_0| \mathcal{M}_0)$\\ [4pt]
  ${ \mathcal{M}_1}$ & $p_1^{C}=p(BF_{01}<1/k_1| \mathcal{M}_1)$  &   $p_1^{M}=p(BF_{01}>k_0| \mathcal{M}_1)$ &  $p_1^{U}=p(1/k_1<BF_{01}<k_0| \mathcal{M}_1)$\\ [4pt]
 \emph{overall} & $p^{C}=\pi_0 \cdot p_0^{C}+ \pi_1 \cdot p_1^{C}$  & $p^{M}=\pi_0 \cdot p_0^{M}+ \pi_1 \cdot p_1^{M}$ & $p^{U}=\pi_0 \cdot p_0^{U}+ \pi_1 \cdot p_1^{U}$\\
 & & & \\
 \hline
\hline \\
\end{tabular}}

}
\label{table:1}
\end{table}

\subsection{Conditional approach to sample size determination} \label{subsec:cond}
\subsubsection{The conditional criterion for sample size determination} 
\label{subsubsec:cond_crit}

In the \emph{conditional} approach to SSD  
the usual  requirement is to guarantee a desired  level of  power (typically 0.8 or higher) at a given value of the parameter of interest. In the Bayesian framework,  
the fixed value is replaced by an entire distribution, namely the design prior, leading to an  \emph{unconditional} power (yet depending on the prior itself).  
Using the BF as a  measure of evidence,  \cite{weiss1997bayesian} considered  SSD for hypothesis testing  based on  type I error rate as well as conditional and unconditional power (without  distinguishing   between analysis and  design prior which had not yet been developed at the time).  Alternative criteria to select a sample size in order to achieve separation of two models with a reasonable \emph{a priori} guarantee were presented in \cite{wang2002simulation}.
 
In the setup of multisite replications, which is the focus of this work, detection of the presence of  heterogeneity (model $\mathcal{M}_1$) is of primary importance. Accordingly we single out $p_1^C$, that is
 the probability of correctly identifying $\mathcal{M}_1$, as most relevant. This leads to the following design criterion  for the optimal selection of  the number of subjects $n^*$ for a  given number of sites $m$:
\begin{align}
\label{eq:n}
n^*    = \min \{ n \in \mathbb{N}: p_1^{C} \geq (1-\beta) \;\; and \;\; p_0^M = \alpha \},
\end{align}
where $\mathbb{N}$ is the set of natural numbers, $(1-\beta)$ is the unconditional power of successfully detecting heterogeneity, while $\alpha$ is the Type I error rate. 

In practice to obtain $n^*$ for a fixed value $m$   one can proceed as follows.   Fix $m$, which is omitted for simplicity from our notation,  set  $n=n_0$  and let $1/k_1^\alpha(n_0)$ be equal to the $\alpha$-quantile of the distribution of $BF_{01}(n_0)$ under  $\mathcal{M}_0$, so that    $p_0^M(n_0) =p(BF_{01}(n_0)<1/k_1^\alpha(n_0) |\mathcal{M}_0)= \alpha$. 
Next evaluate  
$p(BF_{01}(n_0)<1/k_1^\alpha(n_0)  |\mathcal{M}_1)$; if this value is less than $(1-\beta)$, increase $n$ to $n_1>n_0$. 
It appears from Figure \ref{fig:evolution_BF} that the distribution of 
$BF_{01}(n)$ under 
$\mathcal{M}_0$ 
shifts to the right as $n$ increases (for fixed $m$).
Informally we can then conclude that  $k_1^\alpha(n_1)<k_1^\alpha(n_0)$ in order to satisfy the constraint   
$p_0^M(n_1) =p(BF_{01}(n_1)<1/k_1^\alpha(n_1) |\mathcal{M}_0)= \alpha$.
As a consequence, 
$p(BF_{01}(n_1)<1/k_1^\alpha(n_1)  |\mathcal{M}_1) > p(BF_{01}(n_0)<1/k_1^\alpha(n_0)  |\mathcal{M}_1)$
because the distribution of   
$BF_{01}(n)$  under $\mathcal{M}_1$
shifts to the left as $n$ increases.
If $p(BF_{01}(n_0)<1/k_1^\alpha(n_0)  |\mathcal{M}_1)>(1-\beta)$, simply choose $n_1<n_0$.
The optimal solution $n^*$ can then be obtained
\black 
\emph{via} the \emph{Regula Falsi} (False Position)  method  (\citealt{burden2015numerical}), suitably modified to account for the discreteness of $n$.  \texttt{R}-code to evaluate $n^*$ in \eqref{eq:n} is available in the Supplementary material.

\subsubsection{Results {for the conditional approach}}
\label{subsubsec:cond_results}

\begin{center}
\begin{figure}
\includegraphics[width=1 \textwidth]{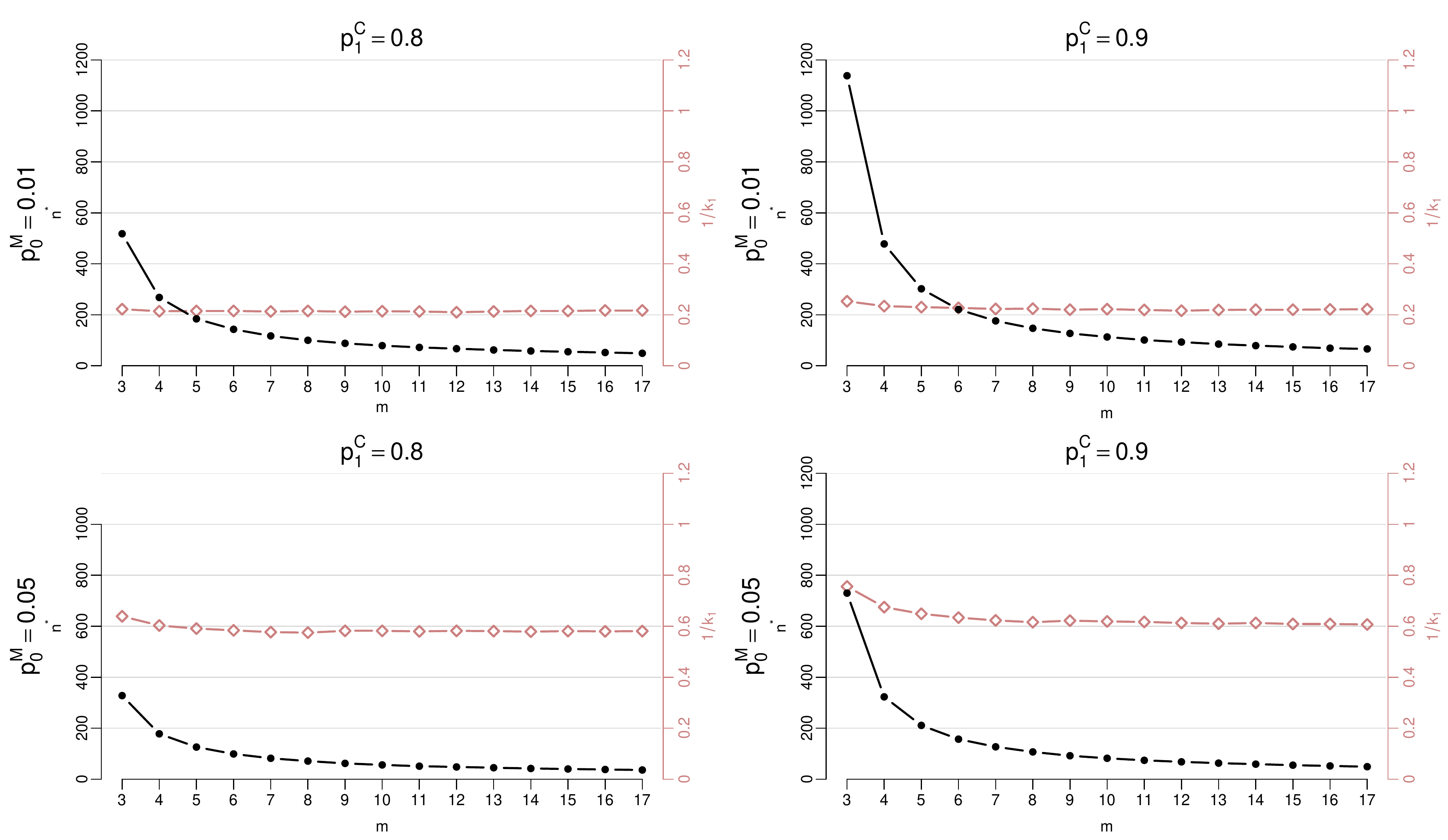}
\centering
\vspace{-1cm}
\caption{Pairs $(m,n^{*})$ (solid circles) for $p_1^{C}\in \{ 0.8, 0.9\}$ and $p_0^{M}\in \{ 0.01, 0.05\}$, along with  corresponding thresholds $1/k_1$ (empty diamonds) {in the conditional approach}.} 
 \label{fig:sim_plot}
\end{figure}
\end{center}

In this subsection we implement the \emph{conditional} approach under four scenarios, corresponding to the  combinations of  
$p_1^{C} \in \{0.80, 0.90\} $ and $p_0^{M} \in \{ 0.01, 0.05\} $, each analysed for  fifteen possible values of the number of sites $m \in \{ 3,4,..., 17\}$. Since interest centres on variability across sites, we excluded the value $m=2$ which would  also lead to much higher values of $n$; see also Figure \ref{fig:sim_plot}.

\begin{table}
\caption{The collection of pairs $(n^{*},m)$ for $p_1^{C}\in \{ 0.8, 0.9\}$ and $p_0^{M}\in \{ 0.01, 0.05\}$, along with the corresponding thresholds $1/k_1$ {in the conditional approach}.}
\vspace{0.2cm}
\centering 
\resizebox{0.75\textwidth}{!}{
\begin{tabular}{cc|cc|cc|cc}
\hline 
\multicolumn{4}{c|}{$p_1^{C}=0.8$} &  \multicolumn{4}{|c}{$p_1^{C}=0.9$} \\[8pt]
\hline 
\multicolumn{2}{c|}{$p_0^{M}=0.01$} & \multicolumn{2}{c|}{$p_0^{M}=0.05$}  &
\multicolumn{2}{|c}{$p_0^{M}=0.01$} & \multicolumn{2}{|c}{$p_0^{M}=0.05$}  \\[6pt]
\hline 
$(n^*,m)$ & $1/k_1$ &  $(n^*,m)$ & $1/k_1$ & $(n^*,m)$ & $1/k_1$ & $(n^*,m)$ & $1/k_1$ \\[1pt]
\hline 
    $(518,3)$ & $0.220$ & $(328,3)$ & $0.639$  & $(1138,3)$ & $0.253$ & $(730,3)$ & $0.756$ \\ [0.1pt]

    $(268,4)$ & $0.214$ & $(178,4)$ & $0.603$  & $(478,4)$ & $0.234$ & $(323,4)$ & $0.675$ \\ [0.1pt]

     $(184,5)$ & $0.215$ &  $(126,5)$ & $0.591$  & $(302,5)$ & $0.230$  & $(211,5)$ & $0.649$\\ [0.1pt]

     $(143,6)$ & $0.215$  & $(99,6)$ & $0.584$  & $(221,6)$ & $0.227$  & $(157,6)$ & $0.634$\\ [0.1pt]

     $(117,7)$ & $0.213$  & $(82,7)$ & $0.577$  & $(176,7)$ & $0.223$  & $(127,7)$ & $0.623$\\ [0.1pt]

     $(100,8)$ & $0.215$ & $(71,8)$ & $0.575$  & $(147,8)$ & $0.224$  & $(107,8)$ & $0.616$\\ [0.1pt]

     $(88,9)$ & $0.212$ & $(62,9)$ & $0.582$  & $(127,9)$ & $0.220$  & $(92,9)$ & $0.622$\\ [0.1pt]

     $(79,10)$ & $0.214$ & $(56,10)$ & $0.582$  & $(113,10)$ & $0.222$  & $(82,10)$ & $0.619$\\ [0.1pt]

     $(72,11)$ & $0.213$ & $(51,11)$ & $0.580$ & $(101,11)$ & $0.219$  & $(74,11)$ & $0.617$ \\ [0.1pt]

     $(67,12)$ & $0.210$  & $(48,12)$ & $0.582$  & $(93,12)$ & $0.216$  & $(68,12)$ & $0.613$\\ [0.1pt]

     $(62,13)$ & $0.213$ & $(45,13)$ & $0.581$ & $(85,13)$ & $0.219$  & $(63,13)$ & $0.610$\\ [0.1pt]

     $(58,14)$ & $0.215$ & $(42,14)$ & $0.579$  & $(79,14)$ & $0.220$  & $(59,14)$ & $0.613$\\ [0.1pt]

     $(55,15)$ & $0.215$ & $(40,15)$ & $0.581$  & $(74,15)$ & $0.220$ & $(55,15)$ & $0.609$\\ [0.1pt]

     $(52,16)$ & $0.217$ & $(38,16)$ & $0.580$ & $(69,16)$ & $0.221$ & $(52,16)$ & $0.609$\\ [0.1pt]

     $(49,17)$ & $0.217$ & $(36,17)$ & $0.581$  & $(66,17)$ & $0.222$ & $(49,17)$ & $0.607$\\ [0.1pt]
\hline 
\hline
\end{tabular}}
\label{table:2}
\end{table}

Priors for $\mu$  and $\gamma$ were chosen as described in Section \ref{Prior}; additionally  we set  $S=10,000$ and $T=50,000$ in Algorithm 1 and 2. Results are presented in Table 2 where thresholds $1/k_1$ are reported along pairs $(n^*,m)$. The same results can be visually inspected  in Figure \ref{fig:sim_plot}. Clearly, for each $m$, a stronger requirement on the probability of correct identification of heterogeneity $p_1^C$ produces a higher value $n^*$, and the same happens when  $p_0^M$ is lowered. For \emph{given} $p_0^M$ the thresholds $1/k_1$ appear robust to the choice of $p_1^C$ and also across configurations $(n^*,m)$,  reflecting low sensitivity in the  lower tail of the distribution of $BF_{01}$  in the range of values under investigation. Notice that a five-fold reduction of $p_0^M$ from 0.05 to 0.01 leads to an approximate three-fold reduction  in $1/k_1$ (roughly from 0.6 to 0.2). On the scale of  evidence in favour of $\mathcal{M}_1$, i.e. $BF_{01}<1/k_1$, based   on \cite{schonbrodt2018bayes},  this translates to an upgrade  from \lq \lq anecdotal\rq \rq{} $(1/3 < BF_{01}<1)$ to \lq \lq{moderate}\rq \rq{} $(1/10<BF_{01}<1/3)$ evidence. This implies that the higher price inherent in a higher $n^*$ might be worth paying not only to achieve a smaller Type I error rate but also to achieve more convincing evidence to detect heterogeneity when it is actually present.  Another feature worth mentioning is the interplay  between $m$ and $n$. Granted that they are inversely related, it emerges that the decrease in $n$ is particularly steep only for small values of $m$, and becomes  relatively modest thereafter; see Figure \ref{fig:sim_plot}. 
The optimal choice $n^*$ is also sensitive  to the \emph{design} prior. If the latter moves toward zero (i.e. closer to the null model),   higher values for $n^*$ are required.  Specifically, keeping $\nu_\gamma$ and $\sigma_\gamma$ constant, and recalling that $\mu_\gamma=0.2$, we need on average a sample size  four times larger if  we move to $\mu_\gamma=0.1$, whereas the sample size can be   halved  if we set $\mu_\gamma=0.3$. The simulated results are analytically provided in the Supplementary material A.
 
It is instructive to compare our results with those presented in  \citet[Table 2]{hedges2021design}. Notice that in their paper a cost function is introduced so that their results depend on  $c_2/c_1$, the ratio of    per-laboratory cost to  per-subject cost. They consider a grid of five values for 
$c_2/c_1$ and, for each of them,  determine the optimal sample size $(n_O,m_O)$ for selected values of the relative variance heterogeneity ($\tau^2/\omega$ in their notation) which corresponds to our $\gamma^2$. The above is replicated at two levels of (conditional)  power, namely 0.8 and 0.9. To compare their results with ours, we first identified a value  $\gamma_0^2=0.04$ of the relative variance which approximately \lq \lq matches\rq\rq{} our  prior expectation $E(\gamma)=0.2$ in the design prior. Next we computed the collection of optimal sample sizes $(n_{Ok},m_{Ok})$ for each $c_2/c_1=r_k$ with $k=1, \ldots,5$.
For given $m_{Ok}$,  a comparison of our optimal  $n_k^*$  
with $n_{Ok}$ shows that the former  is at most only  10\% higher than the latter; remarkably, this   represents a   small increase  in view of the fact that our analysis  fully incorporates uncertainty on the parameter $\gamma$ through an entire distribution on  $\gamma$. By leting  the variance  of the design prior decrease to zero,  we recover sample size results similar to those which hold under the   conditional power approach. 

\subsection{Unconditional approach to sample size determination} \label{subsec:uncond}
\subsubsection{The unconditional criterion for sample size determination} 
\label{subsubsec:uncond_crit}
In the \emph{unconditional} approach to SSD, we consider the overall probabilities $p^{C}$ and $p^{M}$
defined in Table 1.
In this way we modify the design criterion \eqref{eq:n} leading to $n^*$  by replacing the conditional probabilities $p_1^{C}$ and $p_0^M$, with their respective overall probabilities $p^{C}$ and $p^M$. Using the notation of Subsection \ref{subsubsec:cond_crit}, given $(1-\beta)$, $\alpha$ and the  number of sites $m$, the optimal sample size $n^{*}$ is defined following \eqref{eq:n}  as
\begin{align}
\label{eq:n_uncond}
n^{*} = \min \{ n \in \mathbb{N}: p^{C} \geq (1-\beta) \;\; and \;\; p^M = \alpha \},  
\end{align}
where $\mathbb{N}$ is the set of natural numbers, and for simplicity  
we set $p^M_0=p^M_1=\alpha$.
Note that now, differently from the conditional approach,  prior  model probabilities 
are  required.  
For the results in Table 3 
  we additionally fixed  $\pi_0=\pi_1=0.5$. 
  Applying calculations similar to those used for the conditional approach,  we obtain  $1/k_1$, and moreover we  derive $k_0$ as the $(1-\alpha)-$quantile  of the distribution of $BF_{01}$ conditionally on  $\mathcal{M}_1$, under the constraint $p_1^M=\alpha$.
\black 
Once again the effective calculation of $n^*$ is carried out \emph{via} the \emph{Regula Falsi} method, as in  the conditional approach.
 
\subsubsection{Results for the unconditional approach}
\label{subsubsection:uncond_results}

\begin{center}
\begin{figure}
\includegraphics[width=1 \textwidth]{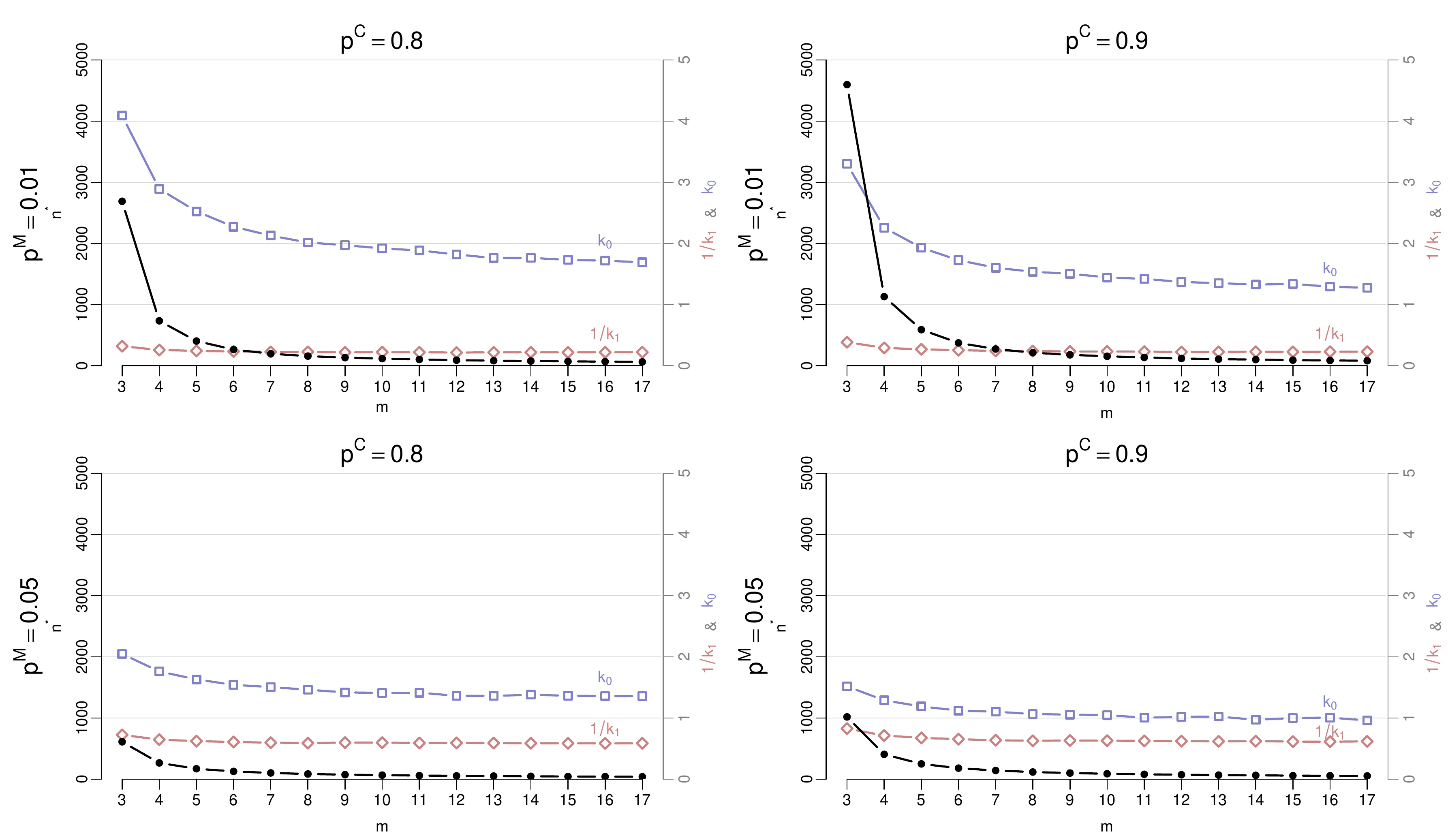}
\centering
\vspace{-1cm}
\caption{Pairs $(m,n^{*})$ (solid circles) for $p^{C}\in \{ 0.8, 0.9\}$ and $p^{M}\in \{ 0.01, 0.05\}$, along with corresponding thresholds $1/k_1$ and $k_0$ (empty diamonds and empty squares respectively) {in the unconditional approach}. } 
 \label{fig:plot_hol2}
\end{figure}
\end{center}

We implement the \emph{unconditional} approach under the combinations of  
$p^{C} \in \{0.80,$ $0.90\} $ and $p^{M} \in \{ 0.01, 0.05\} $, while keeping unchanged the other settings reported in Subsection \ref{subsubsec:cond_results}. The results are tabulated in Table 3 
and graphically represented in Figure \ref{fig:plot_hol2}.

\begin{table}
  \caption{The collection of pairs $(n^{*},m)$ for $p^{C} \in \{0.80,$ $0.90\} $ and $p^{M} \in \{ 0.01, 0.05\} $ along with the decision thresholds $1/k_1$ and $k_0$ {in the unconditional approach}.}
  \vspace{0.2cm}
\label{tab:hol2}
\centering
\resizebox{\textwidth}{!}{
\begin{tabular}{ccc|ccc|ccc|ccc}
\hline 
\multicolumn{6}{c|}{$p^{C}=0.8$} &  \multicolumn{6}{|c}{$p^{C}=0.9$} \\[8pt]
\hline
\multicolumn{3}{c|}{$p^{M}=0.01$} & \multicolumn{3}{c|}{$p^{M}=0.05$}  & \multicolumn{3}{|c}{$p^{M}=0.01$} & \multicolumn{3}{|c}{$p^{M}=0.05$}  \\[6pt]
\hline
   $(n^{*},m)$ & $1/k_1$ & $k_0$ & $(n^{*},m)$ & $1/k_1$ & $k_0$ & $(n^{*},m)$ & $1/k_1$ & $k_0$ & $(n^{*},m)$ & $1/k_1$ & $k_0$ \\ [1pt]
\hline 

   $(2689,3)$ & $0.321$ & $4.090$ & $(611,3)$ & $0.724$ & $2.049$ & $(4596,3)$ & $0.386$ & $3.301$ & $(1018,3)$ & $0.827$ & $1.518$ \\ [0.1pt]

   $(735,4)$ & $0.258$ & $2.891$ & $(266,4)$ & $0.647$ & $1.761$ & $(1130,4)$ & $0.292$ & $2.257$ & $(406,4)$ & $0.714$ & $1.289$ \\ [0.1pt]
   
     $(404,5)$ & $0.244$ & $2.525$ & $(171,5)$ & $0.622$ & $1.632$ & $(590,5)$ & $0.269$ & $1.930$ & $(250,5)$ & $0.675$ & $1.190$ \\ [0.1pt]
    
     $(265,6)$ & $0.234$ & $2.273$ & $(127,6)$ & $0.609$ & $1.542$ & $(374,6)$ & $0.254$ & $1.726$ & $(180,6)$ & $0.654$ & $1.121$ \\ [0.1pt]

     $(197,7)$ & $0.227$ & $2.128$ & $(102,7)$ & $0.597$ & $1.506$ & $(274,7)$ & $0.244$ & $1.603$ & $(142,7)$ & $0.637$ & $1.103$ \\ [0.1pt]

     $(158,8)$ & $0.227$ & $2.015$ & $(85,8)$ & $0.590$ & $1.463$ &   $(214,8)$ & $0.241$ & $1.537$ & $(117,8)$ & $0.627$ & $1.066$ \\ [0.1pt] 
    
     $(133,9)$ & $0.221$ & $1.972$ & $(74,9)$ & $0.598$ & $1.419$ & $(179,9)$ & $0.234$ & $1.504$ & $(100,9)$ & $0.632$ & $1.055$ \\ [0.1pt]

     $(117,10)$ & $0.223$ & $1.918$ & $(66,10)$ & $0.597$ & $1.411$ & $(155,10)$ & $0.234$ & $1.445$ & $(89,10)$ & $0.629$ & $1.048$ \\ [0.1pt]
    
     $(103,11)$ & $0.220$ & $1.886$ & $(59,11)$ & $0.592$ & $1.413$ & $(136,11)$ & $0.231$ & $1.421$ & $(80,11)$ & $0.626$ & $1.005$ \\ [0.1pt]

     $(91,12)$ & $0.216$ & $1.819$ & $(55,12)$ & $0.593$ & $1.364$ & $(119,12)$ & $0.225$ & $1.370$ & $(73,12)$ & $0.622$ & $1.019$ \\ [0.1pt]
    
     $(83,13)$ & $0.219$ & $1.762$ & $(51,13)$ & $0.591$ & $1.361$ & $(108,13)$ & $0.228$ & $1.351$ & $(67,13)$ & $0.618$ & $1.023$ \\ [0.1pt]

     $(78,14)$ & $0.220$ & $1.764$ & $(47,14)$ & $0.588$ & $1.381$ & $(101,14)$ & $0.229$ & $1.337$ & $(63,14)$ & $0.621$ & $0.972$ \\ [0.1pt]

     $(72,15)$ & $0.219$ & $1.731$ & $(44,15)$ & $0.587$ & $1.364$ & $(92,15)$ & $0.227$ & $1.338$ & $(58,15)$ & $0.615$ & $1.000$ \\ [0.1pt]
    
     $(68,16)$ & $0.221$ & $1.718$ & $(42,16)$ & $0.586$ & $1.360$ & $(87,16)$ & $0.229$ & $1.293$ & $(55,16)$ & $0.613$ & $1.007$ \\ [0.1pt]
    
     $(64,17)$ & $0.222$ & $1.691$ & $(40,17)$ & $0.587$ & $1.358$ & $(82,17)$ & $0.230$ & $1.275$ & $(53,17)$ & $0.618$ & $0.960$ \\ [0.1pt]

\hline
\hline \\

\end{tabular}}
\end{table}

Similarly to the conditional approach, the required sample size is dependent on the predetermined probabilities $p^{C}$ and $p^{M}$, while $m$ and $n$ are again inversely related. Regarding the  scale of evidence implied by the thresholds $k_0$ and $1/k_1$, it lies in the range ``anecdotal'' to ``moderate'' for both models. Due to its stricter nature,  the unconditional approach requires larger sample sizes to reach the desired probability of Correct evidence compared to the conditional approach for each $m$, although  the differences become smaller  as  $m$ increases. It seems that the higher sample size required for the unconditional approach is worth paying because not only do we control the probability of misleading evidence under each of the two models, but we also have a lower bound on the probability of overall correct evidence. The latter however  indirectly provides bounds also on the probability of correct evidence under each of the two models especially when $(1-\beta)$ is high and the two model probabilities are not at the extreme of the range $(0,1)$.

\black 

Sensitivity  of SSD to changes in the  \emph{design} prior  appears more pronounced than in the \emph{conditional} approach. Specifically, relative to the benchmark 
$\mu_\gamma=0.2$, 
setting $\mu_\gamma=0.1$ we need on average a sample size six times larger, whereas setting $\mu_\gamma=0.3$ the sample size is more than halved  (approximately 40\%). Detailed results  are available in   Supplementary material B.

\section{Discussion}\label{sec:conclusions}

In this work we have dealt with multiple replication studies, focusing on the \emph{variation} (heterogeneity) of the effect sizes across sites.  Specifically, we  considered the comparison of two models: one without heterogeneity ($\mathcal{M}_0$)  and another one  incorporating heterogeneity ($\mathcal{M}_1$). 
Within this setting, we developed a Bayesian procedure for sample size determination (SSD) capable of  delivering  compelling evidence.
For the two models under consideration, evidence was defined in terms
of the Bayes Factor (BF), which 
was derived  using a suitably defined \emph{analysis} prior.  
 Our design criterion 
was specified through a conditional, as well as an unconditional, approach.
In the former 
the goal 
is to correctly obtain evidence for $\mathcal{M}_1$ (presence of  
 heterogeneity)    {with} high probability,  while assuring that
  the  Type I error rate based on the BF is  
  kept low.
In the unconditional approach instead,  the aim is  to achieve high probability of  
 correct evidence
\emph{overall}  (i.e. averaged across the two  models), while  keeping the  probability of   misleading evidence low, again overall.
The evaluation of our criterion relies on  
the  prior predictive distribution of the BF which was derived  based on the elicitation of a  
\emph{design} prior  (separate from the analysis prior). 


\black 
The  Bayesian methodology presented in this paper  
 represents a  flexible alternative to frequentist based designs for SSD in replication studies.
 A major feature of our  approach is the incorporation of uncertainty through prior distributions, both at the analysis and the design stage,  which significantly extends the standard practice of conditioning on a fixed  value of the parameter as in conventional power analysis. More generally, we can reap the  advantages inherent in  the use of the BF for evidence assessment as described in  \cite{wagenmakers2016bayesian}, and  in particular the possibility of evaluating the posterior probability both for the presence and the absence  of heterogeneity.   

\black 
 We did not  include cost considerations in our design in order to simplify the exposition and focus on the most relevant aspects of our methodology. They could be however incorporated in our framework in a rather straightforward way. Expressing the total cost $C$ in terms of the \emph{per} subject cost   $c_1$,  and  \emph{per} laboratory cost $c_2$, one can simply select among  all pairs $(n^*,m)$ satisfying the design criterion \eqref{eq:n} or \eqref{eq:n_uncond} (see  Table 2 or  Table 3), that specific pair $(n^\prime,m^\prime)$ which minimises the total cost $C$.

An important issue in Bayesian inference, and hence design, is sensitivity of the results to prior specifications.
We performed sensitivity with respect to the \emph{design} prior for relative heterogeneity.
As expected, the required sample size was inversely related with the location of the relative heterogeneity parameter; details and further results are available in    the Supplementary material.  

Despite careful planning,  once the actual experiment is performed, it may happen  that  the outcome will be able to deliver only   an undetermined evidence (see Table 
1), so that neither hypothesis is supported. 
A natural option at this stage is to plan a follow-up design until compelling evidence in either direction, that is in favour of $\mathcal{M}_0$ or $\mathcal{M}_1$, is reached. This is in the spirit of the Sequential Bayes Factor  described in \cite{schonbrodt2017sequential},  which can be implemented either in the open-ended, or maximal sample size, mode.

\section*{Software}
R-code to reproduce the simulated results in the paper along with R-functions to implement our approach using settings different from those employed in our work are available at \href{https://github.com/bourazaskonstantinos/Bayesian-SSD}{https://github.com/bourazaskonstantinos/Bayesian-SSD}.

\section*{Acknowledgments}
This research was partially supported by UCSC (D1 research grants).

\section*{Disclosure Statement}
The authors report there are no competing interests to declare.

\black 
\bibliographystyle{rss}
\bibliography{ref}

\newpage

\section*{Appendix} \label{appendix}
In this appendix we provide the analytical derivation  of the marginal data distribution under  
$\mathcal{M}_0)$ and $\mathcal{M}_1)$, and then  the Bayes factor.

Consider the random vector  $\bm{t_r}=(t_1, \ldots, t_m)$, with $t_j$
the effect size estimate for site $j$, $j=1, \ldots,m$.
Under model $\mathcal{M}_0$  the data distribution is $t_j | \mu, \sim N(\mu, \sigma^2/n)$, whereas under model $\mathcal{M}_1$ it  is $t_j | \mu, \gamma \sim N(\mu, \sigma^2 (1/n+\gamma^2 ))$, all independently. Setting $Q=n\cdot  \sum_{j=1}^m \left( t_j - \bar{t} \right)^2 / \sigma^2 $, {where} $\bar{t}$ is the sample mean of $\bm{t_r}$, the
marginal data distributions  under $\mathcal{M}_0$ is
\begin{align*}
m_0(\bm{t_r}) & =  \mathlarger{ \int\limits _{-\infty}^{+\infty}} f(\bm{t_r} | \mu, \mathcal{M}_0) \pi(\mu) d\mu = \mathlarger{ \int\limits _{-\infty}^{+\infty}} \left( \dfrac{1}{2\pi\sigma^2 /n} \right) ^{m/2} 
\cdot \exp \left\{  - \frac{ \sum_{i=1}^m(t_i - \mu)^2}{2\sigma^2 / n} \right\}   d \mu \\
 & =   \left( \frac{2\pi\sigma^2 }{n} \right) ^{-m/2} \cdot \exp \left\{  - \frac{ \sum_{i=1}^m(t_i - \bar{t})^2}{2\sigma^2 / n} \right\} \mathlarger{ \int\limits _{-\infty}^{+\infty}} \exp \left\{  - \frac{ \sum_{i=1}^m(\bar{t} - \mu)^2}{2\sigma^2 / n} \right\}  d \mu \\[6pt]
  & =   \left( \frac{2\pi\sigma^2 }{n} \right) ^{-m/2} \cdot \exp \left\{  - \frac{ \sum_{i=1}^m(t_i - \bar{t})^2}{2\sigma^2 / n} \right\} \cdot  \left( \dfrac{2\pi\sigma^2 }{n \cdot m} \right) ^{1/2} \\[24pt]
  & =   \left(\frac{1}{m}\right)^{1/2}  \cdot \left(\frac{2\pi \sigma^2}{n}\right)^{(1-m)/2} \cdot \exp \left\{ -\dfrac{Q}{ 2}  \right\}.
\end{align*}.

On the other hand, the marginal data distributions  under $\mathcal{M}_1$ is
\begin{align*}
m_1(\bm{t_r}) & =  \mathlarger{ \int\limits _{0}^{+\infty} \int\limits _{-\infty}^{+\infty}}  f(\bm{t_r} | \mu,\gamma, \mathcal{M}_1) h_a(\gamma) \pi(\mu)  d\mu d\gamma \\[6pt]
 & =   \mathlarger{ \int\limits _{0}^{+\infty} \int\limits _{-\infty}^{+\infty} } \left( 2\pi\sigma^2 \left(\dfrac{1}{n}+\gamma^2\right) \right) ^{-m/2} \cdot \exp  \left\{ - \frac{ n\sum_{i=1}^m(t_i - \mu)^2}{2\sigma^2 \left({1}+n\gamma^2\right)} \right \}  h_a(\gamma) \pi(\mu)  d\mu d\gamma\\
 & =   \mathlarger{ \int\limits _{0}^{+\infty} } \left( 2\pi\sigma^2 \left(\frac{1}{n}+\gamma^2\right) \right) ^{-m/2} \cdot \exp \left\{  - \frac{ n\sum_{i=1}^m(t_i - \bar{t})^2}{2\sigma^2 \left(1+n\gamma^2\right)} \right\} h_a(\gamma) \times \\[6pt]
 & \times  \mathlarger{ \int\limits _{-\infty}^{+\infty}}  \exp \left\{  - \frac{ n\sum_{i=1}^m(\bar{t} - \mu)^2}{2\sigma^2 \left(1+n\gamma^2\right) } \right\}  d \mu d\gamma \\[6pt]
  & =   \mathlarger{ \int\limits _{0}^{+\infty} } \left( 2\pi\sigma^2 \left(\frac{1}{n}+\gamma^2\right) \right) ^{-m/2} \cdot \exp \left\{  - \frac{ n\sum_{i=1}^m(t_i - \bar{t})^2}{2\sigma^2 \left(1+n\gamma^2\right)} \right\}  \\[6pt]
  & \times    \left( \dfrac{2\pi\sigma^2 }{m}  \left(\frac{1}{n}+\gamma^2\right) \right) ^{1/2} h_a(\gamma) d\gamma \\[6pt]
  & =   \left(\frac{1}{m}\right)^{1/2}  \cdot \left(2\pi \sigma^2\right)^{(1-m)/2} \mathlarger{ \int\limits _{0}^{+\infty}} \left(\frac{1}{n}+\gamma^2\right)^{(1-m)/2} \exp \left\{ -\dfrac{Q}{ 2} \cdot  \left( 1+n\gamma^2\right)^{-1}  \right\} h_a(\gamma) d\gamma.
\end{align*}

The resulting BF is:
\vspace{-0.1cm}
\begin{align*}
BF_{01}(\bm{t_r}) & =  \frac{m_0(\bm{t_r})}{m_1(\bm{t_r})} = \frac{n^{(m-1)/2} \cdot \exp \left\{ -\dfrac{Q}{ 2}  \right\}}{\mathlarger{ \int\limits _{0}^{+\infty}} \left(\dfrac{1}{n}+\gamma^2\right)^{(1-m)/2} \exp \left\{ -\dfrac{Q}{ 2} \cdot  \left( 1+n\gamma^2\right)^{-1}  \right\} h_a(\gamma) d\gamma}.
\end{align*}

\section*{Supplementary material}

\subsection*{A. Design prior sensitivity analysis for the conditional approach}
\begin{figure}
	\centering
	\includegraphics[width=1 \textwidth]{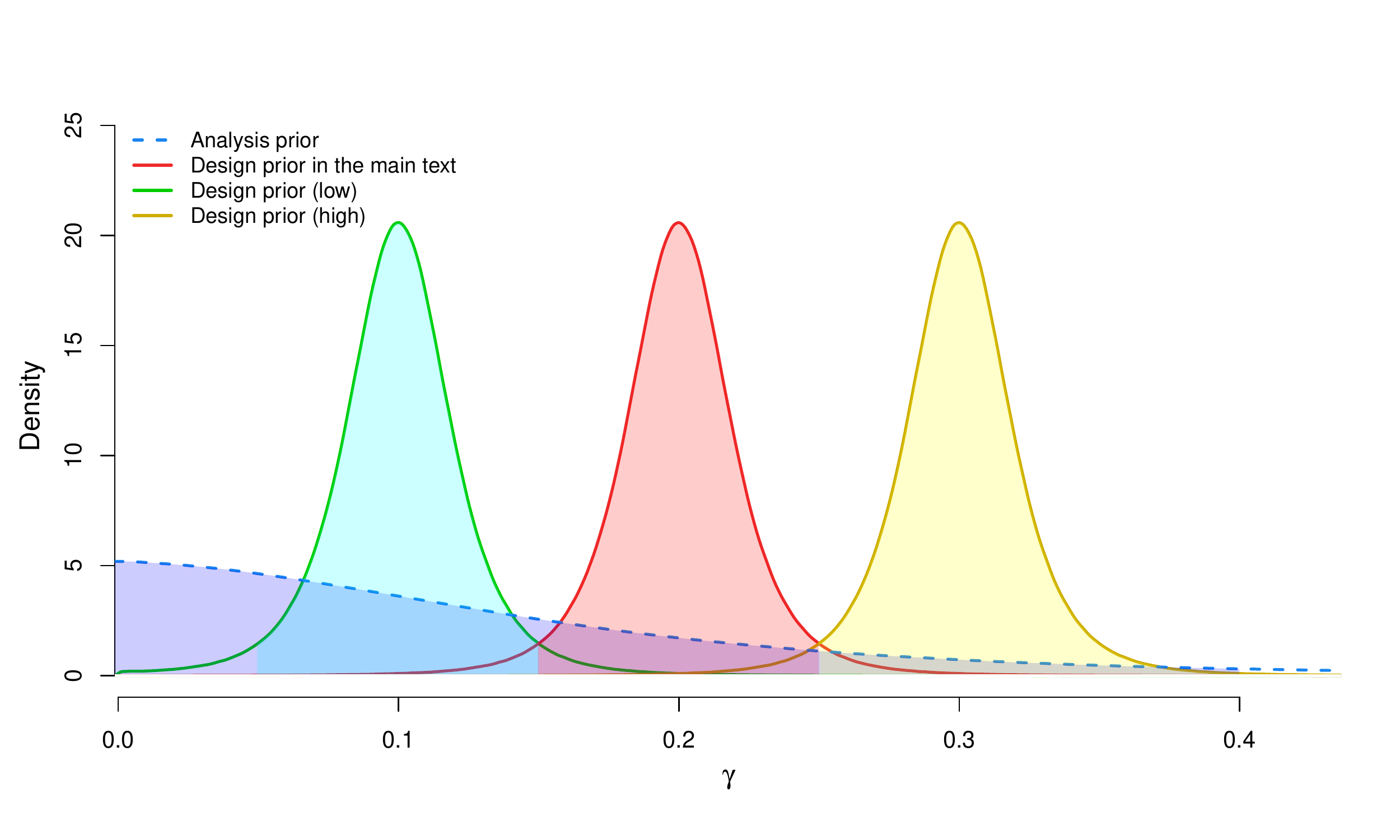}
	\centering
	\caption{The analysis prior $h_a(\gamma)$ (dotted line) and three \emph{design} priors $h_d(\gamma)$ (solid line), along with their $95\%$ credible intervals  (highlighted regions).\\
	}
	\label{fig:sensitivity_plot}
\end{figure}

We perform  sensitivity analysis with respect to the design prior  for \emph{conditional} approach proposed in Subsection 3.2. Specifically, we consider two additional \emph{design} priors, which differ from that in the main text only for the location hyper-parameter $\mu_\gamma$. In the first distribution, we set $\mu_\gamma=0.1$, resulting in the 95\%  credible interval  $[0.05,0.15]$. From a practical point of view, this new prior is located at lower values of $\mu_\gamma$, representing a sensitive set-up, where we are interested to detect a \emph{relative heterogeneity} of small size. On the other hand, the second design prior is located at higher values of $\mu_\gamma$, representing a more tolerant level of heterogeneity. Precisely, we set $\mu_\gamma=0.3$ and this led  to the  95\% credible interval  $[0.25,0.35]$. Figure \ref{fig:sensitivity_plot} provides the density plots for the two additional priors, along with that of the design prior in the main text ($\mu_\gamma=0.2$) and the analysis prior.

\begin{center}
		\begin{figure}
			\includegraphics[width=1 \textwidth]{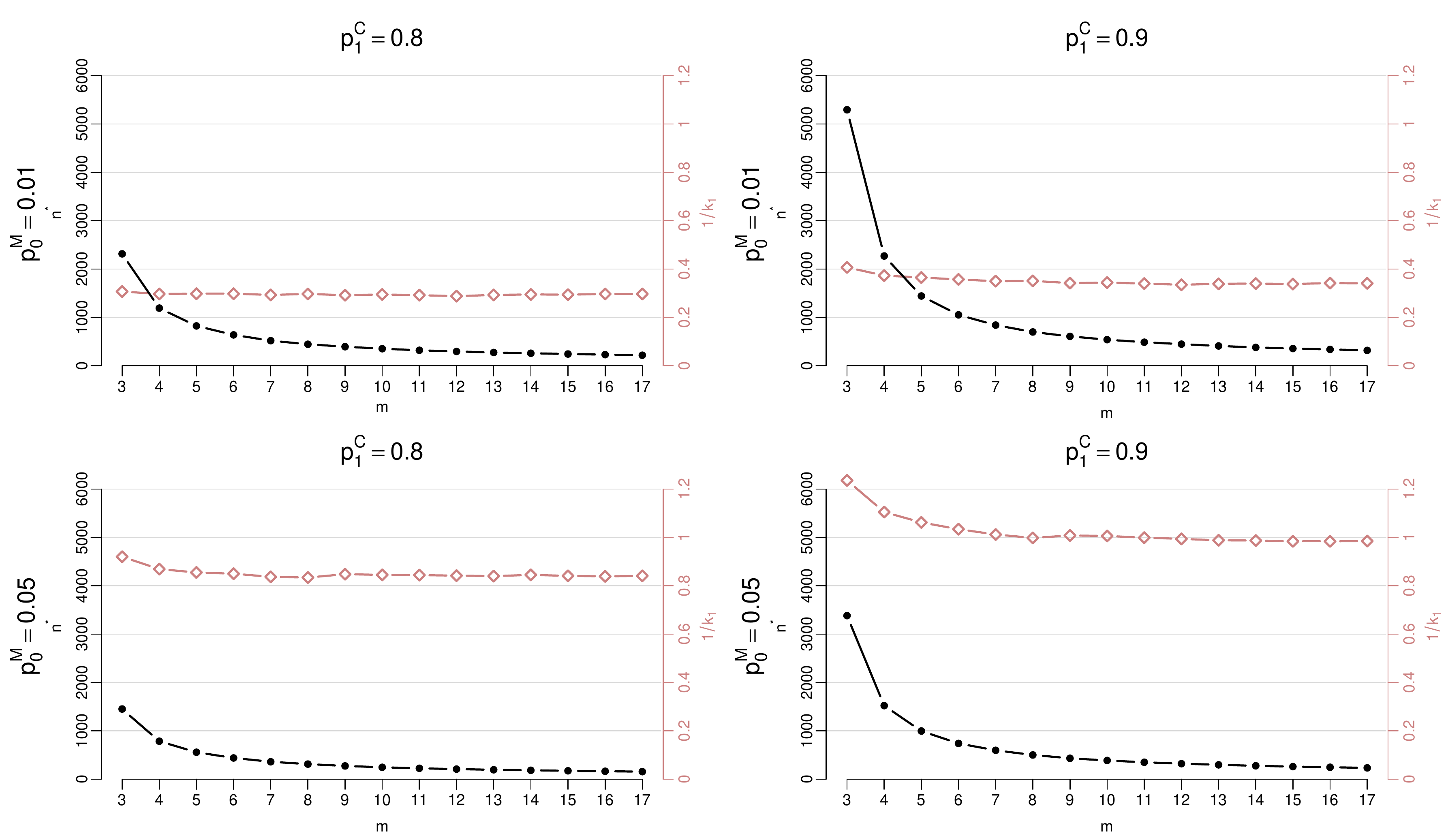}
			\centering
			\caption{Pairs $(m,n^{*})$ (solid circles) for $p_1^{C}\in \{ 0.8, 0.9\}$ and $p_0^{M}\in \{ 0.01, 0.05\}$, along with  corresponding thresholds $1/k_1$ (empty diamonds) {for the conditional approach}. The \emph{design} prior used is Folded-$t(\nu=4,\;\mu_\gamma=0.1,\; \sigma_\gamma=1/55)$.} 
			\label{fig:plot_low}
		\end{figure}
	\end{center}

\begin{center}
		\begin{figure}
			\includegraphics[width=1 \textwidth]{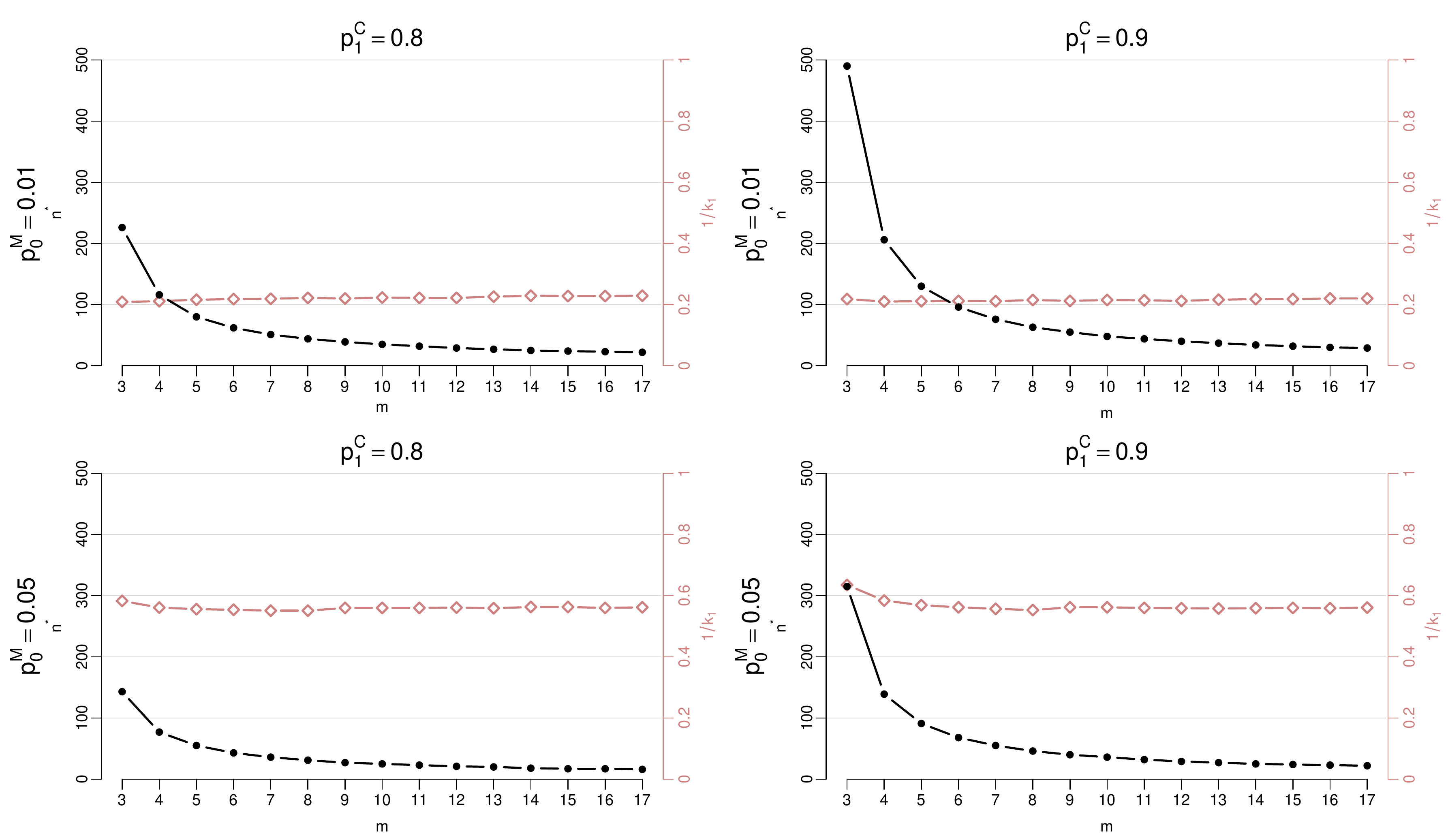}
			\centering
			\caption{Pairs $(m,n^{*})$ (solid circles) for $p_1^{C}\in \{ 0.8, 0.9\}$ and $p_0^{M}\in \{ 0.01, 0.05\}$, along with  corresponding thresholds $1/k_1$ (empty diamonds) {for the conditional approach}. The \emph{design} prior used is Folded-$t(\nu=4,\;\mu_\gamma=0.3,\; \sigma_\gamma=1/55)$.} 
			\label{fig:plot_high}
		\end{figure}
	\end{center}

Tables 1 
and 2 
provide the results of the sensitivity analysis, while the corresponding graphs are in Figures \ref{fig:plot_low} and \ref{fig:plot_high}. As expected, when the design prior 
for the \emph{ relative heterogeneity} $\gamma$
shifts to lower values,   larger sample sizes are required, while the opposite occurs when the distribution shifts to higher values of $\gamma$.
	On the other hand, the thresholds $1/k_1$ on the scale of evidence in favour of $\mathcal{M}_1$ do not change across   the design priors, i.e. ``anecdotal'' for $p_0^M=0.05$ and ``moderate'' for $p_0^M=0.01$. 
	\black 
	
	\black 
	\begin{table}
		\caption{The collection of pairs $(n^{*},m)$ for $p_1^{C}\in \{ 0.8, 0.9\}$ and $p_0^{M}\in \{ 0.01, 0.05\}$, along with the corresponding thresholds $1/k_1$ {for the conditional approach}. The \emph{design} prior used is Folded-$t(\nu=4,\;\mu_\gamma=0.1,\; \sigma_\gamma=1/55)$.} 
		\label{tab:res_low}
		\centering {
			\begin{tabular}{cc|cc|cc|cc}
				\hline 
				\multicolumn{4}{c|}{$p_1^{C}=0.8$} &  \multicolumn{4}{|c}{$p_1^{C}=0.9$} \\[8pt]
				\hline 
				\multicolumn{2}{c|}{$p_0^{M}=0.01$} & \multicolumn{2}{c|}{$p_0^{M}=0.05$}  &
				\multicolumn{2}{|c}{$p_0^{M}=0.01$} & \multicolumn{2}{|c}{$p_0^{M}=0.05$}  \\[6pt]
				\hline 
				$(n^*,m)$ & $1/k_1$ &  $(n^*,m)$ & $1/k_1$ & $(n^*,m)$ & $1/k_1$ & $(n^*,m)$ & $1/k_1$ \\[1pt]
				\hline 
				$(2314,3)$ & $0.307$ & $(1451,3)$ & $0.920$  & $(5294,3)$ & $0.407$ & $(3382,3)$ & $1.236$ \\ [0.1pt]
				
				$(1192,4)$ & $0.297$ & $(785,4)$ & $0.869$  & $(2270,4)$ & $0.373$ & $(1521,4)$ & $1.105$ \\ [0.1pt]
				
				$(824,5)$ & $0.298$ &  $(557,5)$ & $0.855$  & $(1443,5)$ & $0.365$  & $(996,5)$ & $1.062$\\ [0.1pt]
				
				$(639,6)$ & $0.298$  & $(439,6)$ & $0.850$  & $(1053,6)$ & $0.357$  & $(739,6)$ & $1.034$\\ [0.1pt]
				
				$(519,7)$ & $0.293$  & $(360,7)$ & $0.837$  & $(841,7)$ & $0.350$  & $(597,7)$ & $1.012$\\ [0.1pt]
				
				$(445,8)$ & $0.297$ & $(312,8)$ & $0.834$  & $(699,8)$ & $0.351$  & $(501,8)$ & $0.998$\\ [0.1pt]
				
				$(394,9)$ & $0.292$ & $(274,9)$ & $0.848$  & $(608,9)$ & $0.342$  & $(432,9)$ & $1.008$\\ [0.1pt]
				
				$(353,10)$ & $0.295$ & $(246,10)$ & $0.845$  & $(540,10)$ & $0.344$  & $(387,10)$ & $1.006$\\ [0.1pt]
				
				$(320,11)$ & $0.292$ & $(225,11)$ & $0.844$ & $(487,11)$ & $0.340$  & $(350,11)$ & $0.999$ \\ [0.1pt]
				
				$(296,12)$ & $0.288$  & $(208,12)$ & $0.842$  & $(447,12)$ & $0.335$  & $(321,12)$ & $0.994$\\ [0.1pt]
				
				$(275,13)$ & $0.293$ & $(195,13)$ & $0.840$ & $(410,13)$ & $0.339$  & $(297,13)$ & $0.988$\\ [0.1pt]
				
				$(259,14)$ & $0.295$ & $(184,14)$ & $0.845$  & $(380,14)$ & $0.340$  & $(277,14)$ & $0.987$\\ [0.1pt]
				
				$(243,15)$ & $0.294$ & $(173,15)$ & $0.841$  & $(356,15)$ & $0.338$ & $(260,15)$ & $0.984$\\ [0.1pt]
				
				$(230,16)$ & $0.297$ & $(164,16)$ & $0.839$ & $(338,16)$ & $0.342$ & $(247,16)$ & $0.984$\\ [0.1pt]
				
				$(218,17)$ & $0.297$ & $(156,17)$ & $0.841$  & $(318,17)$ & $0.341$ & $(234,17)$ & $0.985$\\ [0.1pt]
				\hline 
			\end{tabular}
			
		}
		
	\end{table}
	
	\begin{table}
		\caption{The collection of pairs $(n^{*},m)$ for $p_1^{C}\in \{ 0.8, 0.9\}$ and $p_0^{M}\in \{ 0.01, 0.05\}$, along with the corresponding thresholds $1/k_1$ {for the conditional approach}. The \emph{design} prior used is Folded-$t(\nu=4,\;\mu_\gamma=0.3,\; \sigma_\gamma=1/55)$.} 
		\centering {
			\begin{tabular}{cc|cc|cc|cc}
				\hline 
				\multicolumn{4}{c|}{$p_1^{C}=0.8$} &  \multicolumn{4}{|c}{$p_1^{C}=0.9$} \\[8pt]
				\hline 
				\multicolumn{2}{c|}{$p_0^{M}=0.01$} & \multicolumn{2}{c|}{$p_0^{M}=0.05$}  &
				\multicolumn{2}{|c}{$p_0^{M}=0.01$} & \multicolumn{2}{|c}{$p_0^{M}=0.05$}  \\[6pt]
				\hline 
				$(n^*,m)$ & $1/k_1$ &  $(n^*,m)$ & $1/k_1$ & $(n^*,m)$ & $1/k_1$ & $(n^*,m)$ & $1/k_1$ \\[1pt]
				\hline 
				$(226,3)$ & $0.209$ & $(143,3)$ & $0.583$  & $(490,3)$ & $0.218$ & $(315,3)$ & $0.635$ \\ [0.1pt]
				
				$(116,4)$ & $0.211$ & $(77,4)$ & $0.561$  & $(206,4)$ & $0.210$ & $(139,4)$ & $0.584$ \\ [0.1pt]
				
				$(80,5)$ & $0.216$ &  $(55,5)$ & $0.556$  & $(130,5)$ & $0.211$  & $(91,5)$ & $0.569$\\ [0.1pt]
				
				$(62,6)$ & $0.218$  & $(43,6)$ & $0.554$  & $(96,6)$ & $0.212$  & $(68,6)$ & $0.562$\\ [0.1pt]
				
				$(51,7)$ & $0.219$  & $(36,7)$ & $0.551$  & $(76,7)$ & $0.211$  & $(55,7)$ & $ 0.557$\\ [0.1pt]
				
				$(44,8)$ & $0.222$ & $(31,8)$ & $0.551$  & $(63,8)$ & $0.215$  & $(46,8)$ & $0.553$\\ [0.1pt] 
				
				$(39,9)$ & $0.220$ & $(27,9)$ & $0.560$  & $(55,9)$ & $0.212$  & $(40,9)$ & $0.562$\\ [0.1pt]
				
				$(35,10)$ & $0.223$ & $(25,10)$ & $0.560$  & $(48,10)$ & $0.215$  & $(36,10)$ & $0.562$\\ [0.1pt]
				
				$(32,11)$ & $0.222$ & $(23,11)$ & $0.560$ & $(44,11)$ & $0.214$  & $(32,11)$ & $0.560$ \\ [0.1pt]
				
				$(29,12)$ & $0.222$  & $(21,12)$ & $0.561$  & $(40,12)$ & $0.212$  & $(29,12)$ & $0.559$\\ [0.1pt]
				
				$(27,13)$ & $0.226$ & $(20,13)$ & $0.559$ & $(37,13)$ & $0.216$  & $(27,13)$ & $0.558$\\ [0.1pt]
				
				$(25,14)$ & $0.229$ & $(18,14)$ & $0.563$  & $(34,14)$ & $0.218$  & $(25,14)$ & $0.559$\\ [0.1pt]
				
				$(24,15)$ & $0.228$ & $(17,15)$ & $0.563$  & $(32,15)$ & $0.218$ & $(24,15)$ & $0.560$\\ [0.1pt]
				
				$(23,16)$ & $0.228$ & $(17,16)$ & $0.560$ & $(30,16)$ & $0.210$ & $(23,16)$ & $0.559$\\ [0.1pt]
				
				$(22,17)$ & $0.229$ & $(16,17)$ & $0.562$  & $(29,17)$ & $0.220$ & $(22,17)$ & $0.561$\\ [0.1pt]
				\hline 
			\end{tabular}
			
		}
		\label{tab:res_high}
	\end{table}

	\subsection*{B. Design prior sensitivity analysis for the unconditional approach}
\begin{center}
		\begin{figure}
			\includegraphics[width=1 \textwidth]{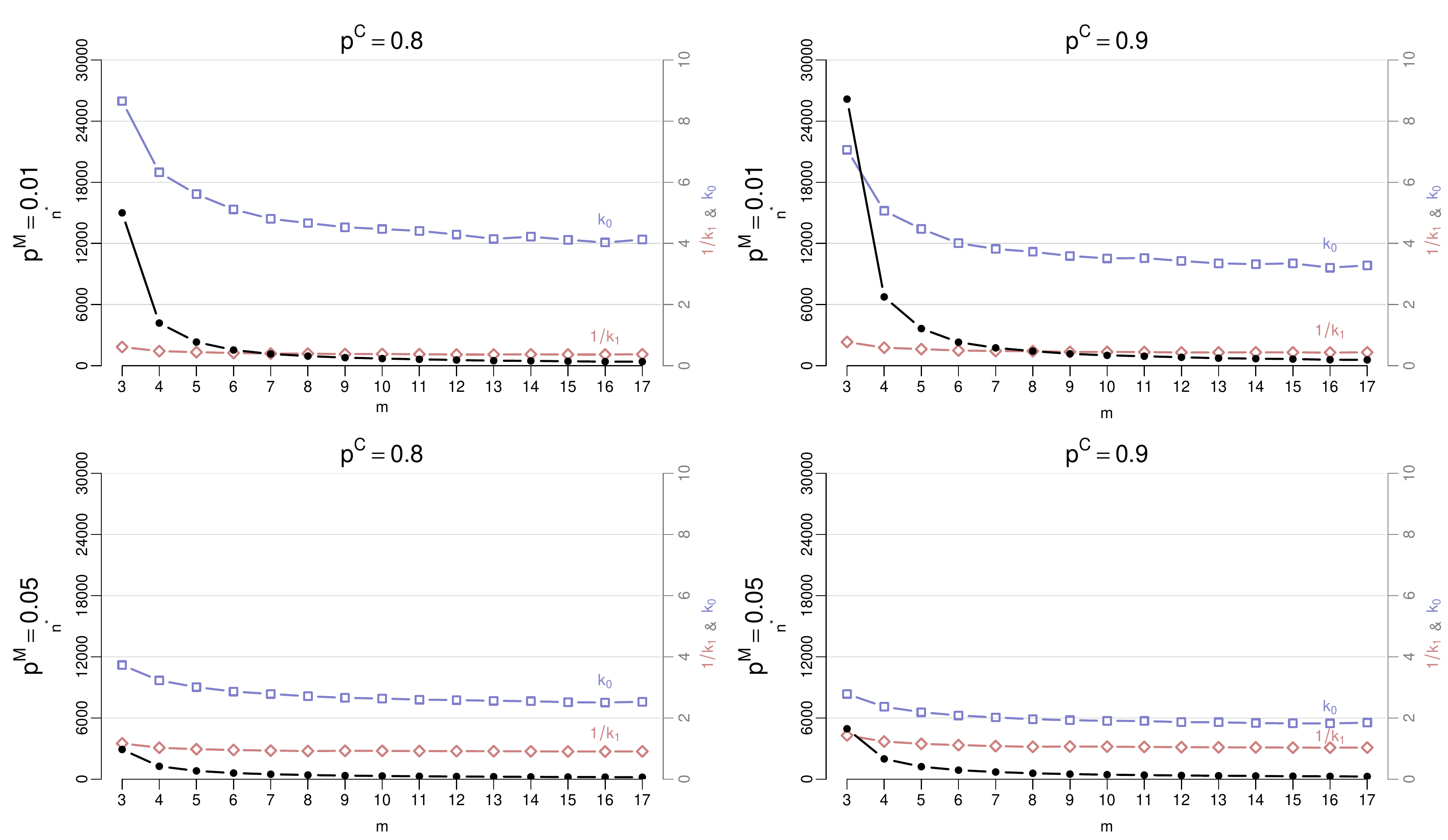}
			\centering
			\caption{Pairs $(m,n^{*})$ (solid circles) for $p^{C}\in \{ 0.8, 0.9\}$ and $p^{M}\in \{ 0.01, 0.05\}$, along with corresponding thresholds $1/k_1$ and $k_0$ (empty diamonds and empty squares respectively) {for the unconditional approach}. The \emph{design} prior used is Folded-$t(\nu=4,\;\mu_\gamma=0.1,\; \sigma_\gamma=1/55)$.} 
			\label{fig:plot_hol1}
		\end{figure}
	\end{center}

\begin{center}
		\begin{figure}
			\includegraphics[width=1 \textwidth]{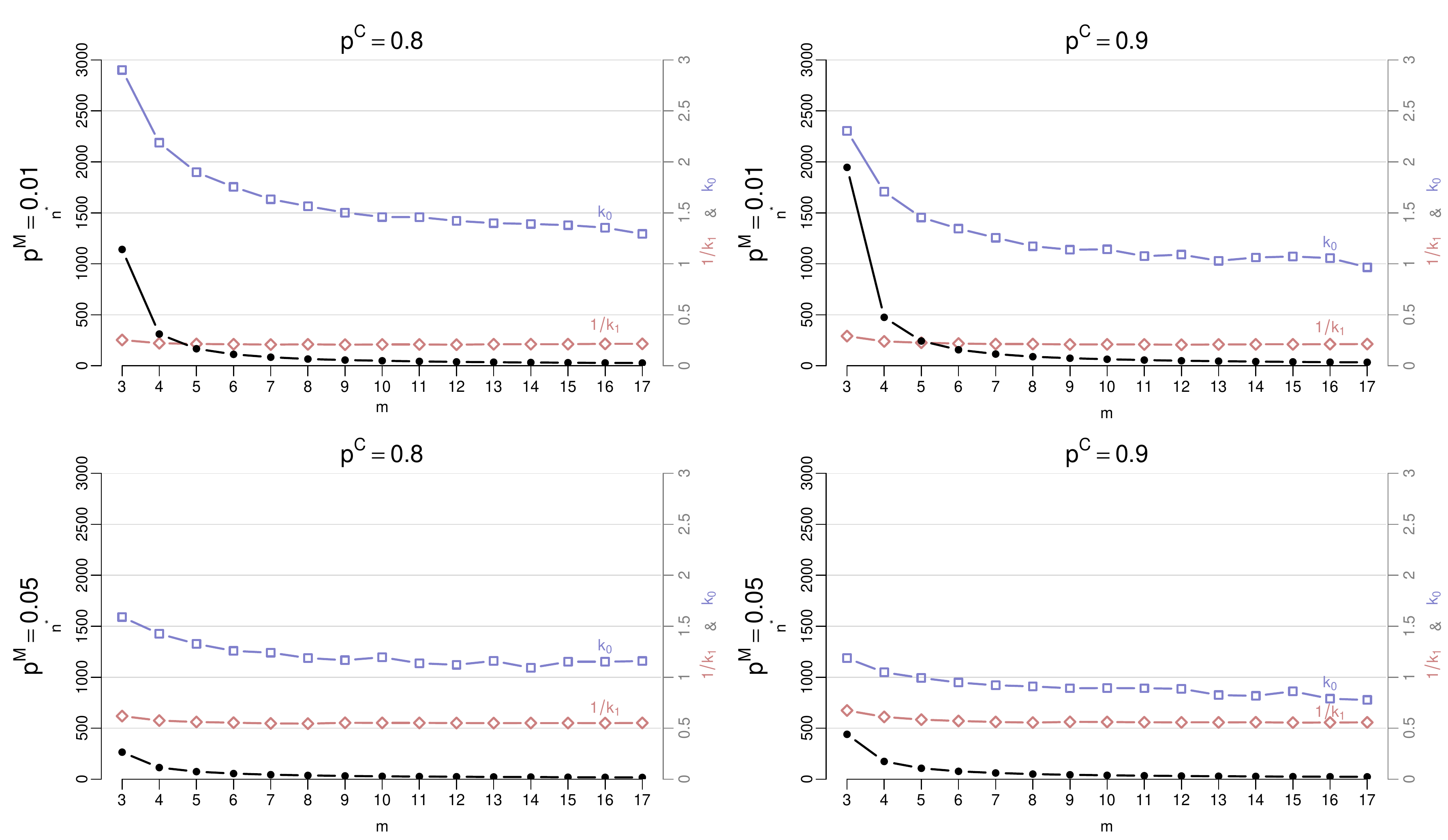}
			\centering
			\caption{Pairs $(m,n^{*})$ (solid circles) for $p^{C}\in \{ 0.8, 0.9\}$ and $p^{M}\in \{ 0.01, 0.05\}$, along with corresponding thresholds $1/k_1$ and $k_0$ (empty diamonds and empty squares respectively) {for the unconditional approach}. The \emph{design} prior used is Folded-$t(\nu=4,\;\mu_\gamma=0.3,\; \sigma_\gamma=1/55)$.} 
			\label{fig:plot_hol3}
		\end{figure}
	\end{center}

	Here, we present the performance of the \emph{unconditional} approach for two additional \emph{design} priors, namely those  introduced in Supplementary material A, i.e. with the \emph{relative heterogeneity} lying in the ranges 5\%-15\% and 25\%-35\%, respectively. The simulated results are tabulated in Tables 3 
	and 4 
	and their graphical representation are reported in Figures \ref{fig:plot_hol1} and \ref{fig:plot_hol3}.
	As in the main text we assumed $p^M_0=p^m_1=\alpha$ and $\pi_0=\pi_1=0.5$.
	
	\black 
	{Comments on the effect of the variation of $\mu_\gamma$ in the design prior are analogous to those reported in the Supplementary material A.} It is worth mentioning that for the unconditional approach the scale of evidence loses its interpretation in some cases. \black Specifically, when the design prior is very close to 0 or too away from it, it is extremely difficult to reach compelling evidence in favour of $\mathcal{M}_1$ or $\mathcal{M}_0$ respectively. Thus, we may obtain values for $1/k_1$ or $k_0$ that are greater or less than 1, respectively. 
	
	\black 
	\begin{table}
		\hspace{-2cm}
		\caption{The collection of pairs $(n^{*},m)$ along with the decision thresholds $1/k_1$ and $k_0$ {for the unconditional approach}. The \emph{design} prior used is  Folded-$t(\nu=4,\;\mu_\gamma=0.1,\; \sigma_\gamma=1/55)$.} 
		
		\label{tab:hol1}
		\centering
		\resizebox{1\textwidth}{!}{
		\begin{tabular}{ccc|ccc|ccc|ccc}
			\hline 
			\multicolumn{6}{c|}{$p^{C}=0.8$} &  \multicolumn{6}{|c}{$p^{C}=0.9$} \\[8pt]
			\hline
			\multicolumn{3}{c|}{$p^{M}=0.01$} & \multicolumn{3}{c|}{$p^{M}=0.05$}  & \multicolumn{3}{|c}{$p^{M}=0.01$} & \multicolumn{3}{|c}{$p^{M}=0.05$}  \\[6pt]
			\hline
			$(n^{*},m)$ & $1/k_1$ & $k_0$ & $(n^{*},m)$ & $1/k_1$ & $k_0$ & $(n^{*},m)$ & $1/k_1$ & $k_0$ & $(n^{*},m)$ & $1/k_1$ & $k_0$ \\ [1pt]
			\hline 
			
			$(15001,3)$ & $0.614$ & $8.654$ & $(2915,3)$ & $1.170$ & $3.730$ & $(26156,3)$ & $0.777$ & $7.061$ & $(4944,3)$ & $1.431$ & $2.785$ \\ [0.1pt]
			
			$(4188,4)$ & $0.479$ & $6.324$ & $(1263,4)$ & $1.028$ & $3.228$ & $(6756,4)$ & $0.591$ & $5.066$ & $(1991,4)$ & $1.231$ & $2.368$ \\ [0.1pt]
			
			$(2326,5)$ & $0.445$ & $5.610$ & $(813,5)$ & $0.981$ & $3.010$ & $(3648,5)$ & $0.544$ & $4.474$ & $(1225,5)$ & $1.156$ & $2.191$ \\ [0.1pt]
			
			$(1545,6)$ & $0.418$ & $5.115$ & $(601,6)$ & $0.954$ & $2.865$ & $(2311,6)$ & $0.500$ & $4.008$ & $(886,6)$ & $1.114$ & $2.084$ \\ [0.1pt]
			
			$(1159,7)$ & $0.399$ & $4.806$ & $(486,7)$ & $0.934$ & $2.786$ & $(1757,7)$ & $0.480$ & $3.820$ & $(702,7)$ & $1.082$ & $2.019$ \\ [0.1pt]
			
			$(950,8)$ & $0.397$ & $4.666$ & $(408,8)$ & $0.920$ & $2.712$ & $(1438,8)$ & $0.477$ & $3.726$ & $(579,8)$ & $1.059$ & $1.961$ \\ [0.1pt] 
			
			$(799,9)$ & $0.382$ & $5.530$ & $(350,9)$ & $0.928$ & $2.661$ & $(1175,9)$ & $0.453$ & $3.591$ & $(498,9)$ & $1.068$ & $1.928$ \\ [0.1pt]
			
			$(706,10)$ & $0.384$ & $4.470$ & $(314,10)$ & $0.926$ & $2.636$ & $(1023,10)$ & $0.452$ & $3.513$ & $(443,10)$ & $1.063$ & $1.909$ \\ [0.1pt]
			
			$(634,11)$ & $0.378$ & $4.410$ & $(284,11)$ & $0.920$ & $2.601$ & $(937,11)$ & $0.449$ & $3.523$ & $(400,11)$ & $1.054$ & $1.902$ \\ [0.1pt]
			
			$(564,12)$ & $0.367$ & $4.292$ & $(262,12)$ & $0.916$ & $2.588$ & $(832,12)$ & $0.435$ & $3.428$ & $(364,12)$ & $1.047$ & $1.865$ \\ [0.1pt]
			
			$(500,13)$ & $0.367$ & $4.148$ & $(243,13)$ & $0.914$ & $2.562$ & $(745,13)$ & $0.436$ & $3.347$ & $(337,13)$ & $1.042$ & $1.861$ \\ [0.1pt]
			
			$(481,14)$ & $0.373$ & $4.221$ & $(226,14)$ & $0.911$ & $2.550$ & $(695,14)$ & $0.438$ & $3.324$ & $(313,14)$ & $1.039$ & $1.834$ \\ [0.1pt]
			
			$(440,15)$ & $0.368$ & $4.119$ & $(211,15)$ & $0.906$ & $2.516$ & $(657,15)$ & $0.437$ & $3.347$ & $(291,15)$ & $1.032$ & $1.825$ \\ [0.1pt]
			
			$(405,16)$ & $0.368$ & $4.031$ & $(200,16)$ & $0.904$ & $2.509$ & $(592,16)$ & $0.432$ & $3.208$ & $(276,16)$ & $1.030$ & $1.822$ \\ [0.1pt]
			
			$(401,17)$ & $0.375$ & $4.126$ & $(192,17)$ & $0.909$ & $2.529$ & $(578,17)$ & $0.438$ & $3.281$ & $(264,17)$ & $1.033$ & $1.848$ \\ [0.1pt]

			\hline
			\hline \\
			
		\end{tabular}}
	\end{table}

	\begin{table}
		\hspace{-3cm}
		\caption{The collection of pairs $(n^{*},m)$ along with the decision thresholds $1/k_1$ and $k_0$ {for the unconditional approach}. The  \emph{design} prior used is Folded-$t(\nu=4,\;\mu_\gamma=0.3,\; \sigma_\gamma=1/55)$.} 
		\label{tab:hol3}
		\centering
		\resizebox{1\textwidth}{!}{
		\begin{tabular}{ccc|ccc|ccc|ccc}
			\hline 
			\multicolumn{6}{c|}{$p^{C}=0.8$} &  \multicolumn{6}{|c}{$p^{C}=0.9$} \\[8pt]
			\hline
			\multicolumn{3}{c|}{$p^{M}=0.01$} & \multicolumn{3}{c|}{$p^{M}=0.05$}  & \multicolumn{3}{|c}{$p^{M}=0.01$} & \multicolumn{3}{|c}{$p^{M}=0.05$}  \\[6pt]
			\hline
			$(n^{*},m)$ & $1/k_1$ & $k_0$ & $(n^{*},m)$ & $1/k_1$ & $k_0$ & $(n^{*},m)$ & $1/k_1$ & $k_0$ & $(n^{*},m)$ & $1/k_1$ & $k_0$ \\ [1pt]
			\hline 
			
			$(1141,3)$ & $0.253$ & $2.902$ & $(265,3)$ & $0.619$ & $1.591$ & $(1946,3)$ & $0.291$ & $2.340$ & $(440,3)$ & $0.674$ & $1.187$ \\ [0.1pt]
			
			$(311,4)$ & $0.221$ & $2.187$ & $(114,4)$ & $0.575$ & $1.410$ & $(475,4)$ & $0.239$ & $1.707$ & $(174,4)$ & $0.611$ & $1.048$ \\ [0.1pt]
			
			$(167,5)$ & $0.214$ & $1.898$ & $(74,5)$ & $0.561$ & $1.327$ & $(243,5)$ & $0.226$ & $1.455$ & $(107,5)$ & $0.584$ & $0.993$ \\ [0.1pt]
			
			$(112,6)$ & $0.211$ & $1.754$ & $(55,6)$ & $0.554$ & $1.258$ & $(157,6)$ & $0.218$ & $1.345$ & $(77,6)$ & $0.571$ & $0.948$ \\ [0.1pt]
			
			$(84,7)$ & $0.208$ & $1.634$ & $(44,7)$ & $0.547$ & $1.241$ & $(115,7)$ & $0.213$ & $1.256$ & $(61,7)$ & $0.561$ & $0.921$ \\ [0.1pt]
			
			$(66,8)$ & $0.210$ & $1.565$ & $(37,8)$ & $0.545$ & $1.188$ & $(89,8)$ & $0.213$ & $1.173$ & $(50,8)$ & $0.555$ & $0.910$ \\ [0.1pt] 
			
			$(56,9)$ & $0.207$ & $1.502$ & $(32,9)$ & $0.553$ & $1.167$ & $(74,9)$ & $0.209$ & $1.138$ & $(43,9)$ & $0.562$ & $0.892$ \\ [0.1pt]
			
			$(49,10)$ & $0.209$ & $1.459$ & $(28,10)$ & $0.552$ & $1.196$ & $(64,10)$ & $0.210$ & $1.142$ & $(38,10)$ & $0.561$ & $0.894$ \\ [0.1pt]
			
			$(43,11)$ & $0.209$ & $1.458$ & $(26,11)$ & $0.552$ & $1.137$ & $(56,11)$ & $0.209$ & $1.075$ & $(34,11)$ & $0.558$ & $0.892$ \\ [0.1pt]
			
			$(38,12)$ & $0.207$ & $1.421$ & $(24,12)$ & $0.551$ & $1.122$ & $(49,12)$ & $0.206$ & $1.090$ & $(31,12)$ & $0.557$ & $0.887$ \\ [0.1pt]
			
			$(35,13)$ & $0.211$ & $1.399$ & $(22,13)$ & $0.550$ & $1.160$ & $(45,13)$ & $0.209$ & $1.030$ & $(29,13)$ & $0.557$ & $0.825$ \\ [0.1pt]
			
			$(32,14)$ & $0.212$ & $1.391$ & $(21,14)$ & $0.551$ & $1.092$ & $(41,14)$ & $0.210$ & $1.062$ & $(27,14)$ & $0.558$ & $0.817$ \\ [0.1pt]
			
			$(30,15)$ & $0.212$ & $1.379$ & $(19,15)$ & $0.551$ & $1.152$ & $(38,15)$ & $0.210$ & $1.072$ & $(25,15)$ & $0.555$ & $0.862$ \\ [0.1pt]
			
			$(28,16)$ & $0.215$ & $1.355$ & $(18,16)$ & $0.550$ & $1.152$ & $(35,16)$ & $0.212$ & $1.056$ & $(24,16)$ & $0.556$ & $0.789$ \\ [0.1pt]
			
			$(27,17)$ & $0.215$ & $1.293$ & $(17,17)$ & $0.552$ & $1.159$ & $(34,17)$ & $0.213$ & $0.966$ & $(23,17)$ & $0.557$ & $0.777$ \\ [0.1pt]

			\hline
			\hline \\
			
		\end{tabular}}
	\end{table}

\end{spacing}

\end{document}